\documentclass[twocolumn,showpacs,preprintnumbers,amsmath,amssymb]{revtex4}

\usepackage{graphicx}% Include figure files
\usepackage{dcolumn}% Align table columns on decimal point
\usepackage{bm}% bold math

\usepackage{xcolor}
\usepackage[outdir=./]{epstopdf}
\usepackage{amsmath}
\usepackage{cancel}
\DeclareMathOperator{\sgn}{sgn}
\usepackage[mathscr]{euscript}

\begin{document}

%\preprint{APS/123-QED}

\title{Maxwell-consistent, symmetry- and energy-preserving solutions\\for ultrashort laser pulse propagation beyond the paraxial approximation}% Force line breaks with \\

\author{P. Gonz{\'a}lez de Alaiza Mart{\'i}nez}
    \email{pedro.gonzalez@u-bordeaux.fr}
\author{G. Duchateau}
\author{B. Chimier}
\author{R. Nuter}
\affiliation{%
Centre Lasers Intenses et Applications, Universit{\'e} de Bordeaux - CNRS - CEA, UMR 5107, F-33405 Talence, France}

\author{I. Thiele}
    \affiliation{Department of Physics, Chalmers University of Technology, SE-412 96 G{\"o}teborg, Sweden}
    
\author{S. Skupin}
	\affiliation{Institut Lumi{\`e}re Mati{\`e}re, Universit{\'e} Lyon - CNRS, UMR 5306, 69622 Villeurbanne, France}
	
\author{V. T. Tikhonchuk}
	\altaffiliation[Also at ]{Centre Lasers Intenses et Applications, Universit{\'e} Bordeaux - CNRS - CEA, UMR 5107, F-33405 Talence, France}
\affiliation{ELI-Beamlines, Institute of Physics, Czech Academy of Sciences, 25241 Doln{\'i} Bre{\v z}any, Czech Republic}

\date{\today}% It is always \today, today,

\begin{abstract}
We analytically and numerically investigate the propagation of ultrashort tightly focused laser pulses in vacuum, with particular emphasis on Hermite-Gaussian and Laguerre-Gaussian modes. We revisit the Lax series approach for forward-propagating linearly-polarized laser pulses, in order to obtain Maxwell-consistent and symmetry-preserving analytical solutions for the propagation of all field components beyond the paraxial approximation in four-dimensional geometry (space and time). We demonstrate that our solution conserves the energy, which is set by the paraxial-level term of the series. The full solution of the wave equation towards which our series converges is calculated in the Fourier space. Three-dimensional numerical simulations of ultrashort tightly-focused pulses validate our analytical development.
\end{abstract}

\pacs{42, 42.25.Bs, 42.55.-f}% PACS, the Physics and Astronomy
\maketitle

%%%%%%%%%%%%%%%%%%%%%%%%%%%%%%%%%%%%%%%%%%%%%%%%%%%%%%%%%%%%

\section{\label{sec:introduction}Introduction}

Spatial and temporal pulse shaping makes the laser a highly versatile tool for a large number of applications such as micromachining and material processing \cite{Davis96, Gamaly06,Patel2017}, Terahertz generation \cite{Stepanov05, Hebling08}, or biological imaging and non-invasive surgeries \cite{Zipfel03, Chung09}. Paraxial approximation, which assumes that the light angular spectrum is sufficiently  narrow, is widely used to study the propagation of laser beams in weak focusing conditions. However, the applications mentioned above usually require tightly focused ultrashort laser pulses. Modeling the propagation of such laser pulses is a complex problem because the deviation from the principal propagation direction is large and the paraxial approximation is no longer valid.

Electromagnetic codes, such as Particle-In-Cell codes \cite{Birdsall1985,Hockney1988} or codes based on the Unidirectional Pulse Propagation Equation \cite{Kolesik2002,Kolesik2004}, are powerful tools for understanding experiments on laser-matter interaction, where laser field components are commonly known in the far field. In these simulations external electromagnetic waves that enter the computational domain are usually prescribed as paraxial modes on the boundaries, which is not adequate for strongly focused ultrashort laser pulses because the initial distortion may be increased in the course of propagation \cite{Couairon2015}, leading eventually to nonphysical fields in the simulation box. Therefore, there is a need to determine analytical solutions of Maxwell equations for tightly focused laser pulses.

Different analytical models, restricted to specific beam shapes or spatial symmetry conditions, have been developed to describe nonparaxial laser beam propagation in several physical contexts, such as perturbative expansions of the wave equation \cite{Esarey1995,Li2016}, the angular spectrum method \cite{Sepke2006}, transformation optics \cite{Fedorov2016} or analytical solutions based on the Helmholtz equation for laser-driven electron acceleration \cite{Marceau2013,Martens2014}. Lax {\it et al.} \cite{Lax1975} proposed a simple method which paved the way to introduce the nonparaxial corrections to a given paraxial solution in more general situations. They demonstrated that the paraxial solution is actually the zeroth-order consistent solution to the Maxwell equations, obtained by expanding the wave equation (in their case, for a Gaussian linearly-polarized vector potential) using a power series in the beam divergence angle.

The nonparaxial perturbative equations proposed by Lax {\it et al.} were subsequently analyzed in more detail, always on the basis of the wave equation applied to the vector potential, by several authors for either Gaussian beams \cite{Davis1979, Couture1981,Sheppard1999,Salamin2007,Seshadri2008} or Hermite-Gaussian and Laguerre-Gaussian beams \cite{Agrawal1979,Agrawal1983,Takenaka1985,Laabs1997}. Later, Porras {\it et al.} \cite{Porras2001,Varin2006} proposed a similar time-domain perturbative approach, based on a different expansion parameter, in order to study the propagation of vectorial few-cycle light pulses. More recently, Favier {\it et al.} took into account spatio-temporal couplings in the wave equation in order to extend Lax perturbative equations to few-cycle pulses \cite{Favier2017}. In the transverse-spatial and temporal Fourier domain, they linked the Lax series with a Taylor expansion of the exact solution of the wave equation, but their proposed high-order corrections hinged on an arbitrary number of integration constants, which were chosen to match some known nonparaxial solutions.

This paper aims at addressing two problems which remain open despite the advances made in the previous works. The first problem is that all the previous approaches solely dealt with the wave equation (in the cited papers, applied to the vector potential) split into a Lax series, and not with the full set of Maxwell equations when calculating high-order corrections. Since each component of the electric, magnetic and vector potential fields verifies the scalar wave equation, we expect to obtain a unique solution to the Maxwell equations whatever the component chosen to calculate high-order corrections. The second problem is that, when calculating high-order terms of the Lax series from the solutions at lower orders, spurious homogeneous solutions that are not compatible with Maxwell equations may be added through integration constants. We demonstrate in this paper that removing those spurious homogeneous solutions, as well as not breaking the existing symmetry between the electric and magnetic fields, implies preserving the laser energy through transverse planes. Conservation of energy is a fundamental physical principle that, to the best of our knowledge, had never been considered before in the context of nonparaxial corrections. Indeed, in previous works these integration constants were determined by making {\it ad hoc} assumptions, not sufficiently supported by the Maxwell equations, on how the nonparaxial corrections should be at the focal point \cite{Couture1981, Takenaka1985} or on the beam axis \cite{Salamin2007}.

In Sec.~\ref{sec:solweq}, our Lax-series-based analytical solution for all electromagnetic field components is presented. Since Maxwell equations are satisfied, each electromagnetic field component verifies the scalar wave equation. By preserving the existing symmetry between the electric and magnetic fields, recursive relations to obtain the terms of our series are given in the Fourier space and the resulting solution is successfully compared with a numerically exact Maxwell solver \cite{Thiele2016}. Provided that there are no evanescent modes in the paraxial-level term, our solution is convergent. We also demonstrate that our solution conserves the total energy through transverse planes, which is set by the paraxial-level term of our series. This solution as presented in Sec.~\ref{sec:convergence} represents an accurate way of injecting ultrashort laser pulses of arbitrary shape in space and time in codes based on the Unidirectional Pulse Propagation Equation and, under the cost of computing inverse Fourier transforms, also in Finite-Difference-Time-Domain electromagnetic codes. In Sec.~\ref{sec:Asymptotic behavior far from focal plane} we calculate the leading term of the asymptotic limit of our Lax-series-based analytical solution far from the focal plane, for both monochromatic beams and ultrashort laser pulses, which results in paraxial-like expressions. These analytical expressions are a baseline for further developments aiming at obtaining an easy and low-computational-cost means of computing the near fields related to those assumed-known paraxial far fields, avoiding the computation of any Fourier transform. Thanks to three-dimensional Maxwell-consistent numerical simulations carried out with the code {\sc arctic}, based on the Yee scheme \cite{Yee1966}, we discuss the adequacy of prescribing ultrashort laser pulses by the leading term of the asymptotic limit at a finite distance from the focal plane. Conclusions and outlooks are drawn in Sec.~\ref{sec:Conclusion and outlooks}.

%%%%%%%%%%%%%%%%%%%%%%%%%%%%%%%%%%%%%%%%%%%%%%%%%%%%%

\section{\label{sec:solweq}Analytical solutions of Maxwell equations}

\subsection{Maxwell equations and their properties}

Maxwell equations in vacuum read as follows:
\begin{equation}\label{eq:divE=0}
{\pmb \nabla}\cdot{\pmb E} = 0,
\end{equation}
\begin{equation}\label{eq:divB=0}
{\pmb \nabla}\cdot{\pmb B} = 0,
\end{equation}
\begin{equation}\label{eq:Faraday}
\partial_t{\pmb B}+ {\pmb \nabla}\times{\pmb E} = 0,
\end{equation}
\begin{equation}\label{eq:Ampere}
\partial_t{\pmb E}- c^2 {\pmb \nabla}\times{\pmb B} = 0,
\end{equation}
where ${\pmb E}$ and ${\pmb B}$ are the electric and magnetic fields, respectively, and $c$ is the speed of light in vacuum. Maxwell's equations are highly symmetrical and they place the electric and magnetic fields on equal footing \cite{Cameron2012}. Indeed, both electric and magnetic fields verify the wave equation:
\begin{equation}\label{eq:waveeqE}
\left( c^{-2}\partial_t^2-{\pmb \nabla}^2   \right){\pmb E} =  {\pmb 0},
\end{equation}
\begin{equation}\label{eq:waveeqB}
\left( c^{-2}\partial_t^2-{\pmb \nabla}^2   \right){\pmb B} = {\pmb 0}.
\end{equation}

Note that in this paper, we formally present our results in vacuum. For monochromatic or narrow-bandwidth pulses, by replacing $c$ by $c/n_0$, where $n_0$ is a constant refractive index, our results generalize to homogeneous dielectric media. Because our solutions are derived in the Fourier space, it would be straightforward to extend it to shorter pulses with linear dispersion.

\subsection{The wave equation}

Throughout this paper, we shall work in Cartesian coordinates ($x$, $y$, $z$), where $x$ is the optical propagation axis (also referred to as longitudinal axis) and $y$ and $z$ are the transverse coordinates. The beam focus position is placed at $x=0$.

We seek solutions of Maxwell equations that are waves propagating along longitudinal axis according to the following Ansatz:
\begin{equation}\label{eq:Ansatz:E}
{\pmb E}(x,y,z,t) = E_0 \left(  \begin{array}{c}
\psi_{E_x}(x,y,z,t) \\
\psi_{E_y}(x,y,z,t) \\
\psi_{E_z}(x,y,z,t) \\
\end{array}   \right) \, {\rm e}^{{\rm i}(k_0 x-\omega_0 t  )},
\end{equation}
\begin{equation}\label{eq:Ansatz:B}
{\pmb B}(x,y,z,t) = \frac{E_0}{c} \left(  \begin{array}{c}
\psi_{B_x}(x,y,z,t) \\
\psi_{B_y}(x,y,z,t) \\
\psi_{B_z}(x,y,z,t) \\
\end{array}   \right) \, {\rm e}^{{\rm i}(k_0 x-\omega_0 t  )},
\end{equation}
where $\omega_0 = 2\pi c/\lambda_0$ is the angular frequency of the laser field, $\lambda_0$ is the wavelength, $k_0=\omega_0/c$ is the wavenumber, $\psi_{E_x}$, $\psi_{E_y}$, $\psi_{E_z}$, $\psi_{B_x}$, $\psi_{B_y}$ and $\psi_{B_z}$ are the spatio-temporal envelopes of $E_x$, $E_y$, $E_z$, $B_x$, $B_y$ and $B_z$, respectively, and $E_0$ represents the electric field amplitude. Note that in this paper we only seek forward-propagating solutions propagating along $x$ axis, as stated by Ans{\"a}tze~(\ref{eq:Ansatz:E})~and~(\ref{eq:Ansatz:B}), although Eqs.~(\ref{eq:waveeqE})~and~(\ref{eq:waveeqB}) admit, in general, bidirectional solutions. Implicitly, we require that ${\pmb E}$ and ${\pmb B}$ have no evanescent components. Moreover, because they are complex fields, the negative frequency components are required to be the complex conjugates of their respective positive frequency components \cite{Berge2007}.

By substituting Eq.~\eqref{eq:Ansatz:E} into Eq.~\eqref{eq:waveeqE} and Eq.~\eqref{eq:Ansatz:B} into Eq.~\eqref{eq:waveeqB}, each of the six spatial envelopes, generically denoted as $\psi$, verifies the so-called wave equation:
\begin{equation}\label{eq:psi prime}
{\pmb \nabla}_{\perp}^2 \psi + 2 {\rm i} k_0 \left[\partial_x + \frac{\partial_t}{c} \right] \psi = - \partial_x^2 \psi + \frac{\partial_t^2\psi}{c^2} ,
\end{equation}
where ${\pmb \nabla}_{\perp}^2 = \partial_y^2+\partial_z^2$. It is useful to express Eq.~\eqref{eq:psi prime} in the laser co-moving reference system $x' = x$ and $t' = t - x/c$:
\begin{equation}\label{eq:psi}
{\pmb \nabla}_{\perp}^2 \psi + 2 {\rm i} k_0 \left(1 +\frac{{\rm i}\partial_{t'}}{\omega_0} \right) \partial_{x'} \psi = - \partial_{x'}^2 \psi .
\end{equation}

The paraxial approximation neglects the term on the right-hand side of Eq.~\eqref{eq:psi} by claiming that the field variation along $x$ axis is small compared to the wavelength $\lambda_0$ (i.e., the wavefront is considered to be almost perpendicular to $x$ axis) and to the transverse variation along $y$ and $z$ axes (i.e., the transverse profile is supposed to remain almost unchanged over a distance of the order of $\lambda_0$). Considering $D_0$ the $1/{\rm e}$ diameter of the Gaussian solution at the beam focus (we assume that the diameter is the same along $y$ and $z$ axis) and $x_R = \pi  D_0^2 / (4 \lambda_0)$ the associated Rayleigh length, we reformulate Eq.~\eqref{eq:psi} in the dimensionless coordinates $\xi = x' / x_R $, $\tau = \omega_0 t'$, $\upsilon =2 y / D_0$ and $\zeta = 2 z /D_0$ as follows:
\begin{equation}\label{eq:psi:xiupsilonzeta}
\partial_\perp^2\psi + 4{\rm i} \, T \, \partial_\xi \psi = - \varepsilon^2 \partial_\xi^2 \psi,
\end{equation}
where $\partial_\perp^2=\partial_\upsilon^2+\partial_\zeta^2$ and the operator $T= 1+{\rm i}\,\partial_\tau$ introduces the space-time focusing \cite{Brabec1997,Berge2007}. Equation~\eqref{eq:psi:xiupsilonzeta} reveals that the term on the right-hand side is actually a small correction of order of $\varepsilon^2$, where $\varepsilon =  D_0 / (2 x_R)$ is the tangent of the beam divergence angle and is assumed to be small in the paraxial limit. For arbitrary spatial beam shapes, for which the Gaussian 1/e~beam diameter $D_0$ does not apply, one can define $\varepsilon$ as the angular spectral width divided by $k_0$. Note that monochromatic solutions are given by Eq.~\eqref{eq:psi:xiupsilonzeta} in the limit $T\rightarrow 1$, which means that the time variation of the envelopes is negligible.

Equation~(\ref{eq:psi:xiupsilonzeta}) in the transverse-spatial and temporal Fourier domain (see Appendix~\ref{sec:sup_mat_Fourier}) reads:
\begin{equation}\label{eq:psi:xiupsilonzeta:Fourier}
\left( \frac{{\rm i}\kappa_\perp^2}{4 \, \hat{T}}  +  \partial_\xi - \frac{{\rm i}\varepsilon^2}{4 \, \hat{T}} \partial_\xi^2   \right) \hat{\psi} = 0,
\end{equation}
where $\kappa_\perp^2 = \kappa_y^2+\kappa_z^2$, $\kappa_y = D_0 k_y /2$, $\kappa_z = D_0 k_z /2$, $\hat{T} = 1 + \Omega$, and $\Omega = \omega/\omega_0$. Restricting the temporal bandwidth of the complex fields $\pmb E$ and $\pmb B$ to the positive frequency range implies that $\Omega \ll 1$. The exact forward-propagating solution of Eq.~\eqref{eq:psi:xiupsilonzeta:Fourier}, with the boundary condition placed at $\xi=0$, reads:
\begin{equation}\label{eq:Illia solver:psi}
\begin{split}
&\hat{\psi}(\xi,\kappa_y,\kappa_z,\Omega) =\\
& \hat{\psi}(0,\kappa_y,\kappa_z,\Omega) \; {\rm e}^{-\frac{2{\rm i}\, \hat{T}}{\varepsilon^2} \left(1 - \sqrt{1-\frac{\varepsilon^2\kappa_\perp^2}{4\, \hat{T}^2}}\right)\xi},
\end{split}
\end{equation}
which, by abuse of language, will be called {\it general solution of the wave equation} all through this paper in spite of its lack of bidirectionality.

Equation~\eqref{eq:Illia solver:psi} discloses that the exact forward-propagating solution preserves its complex module in all transverse planes:
\begin{equation}\label{eq:Illia solver:psi:mod}
| \hat{\psi}(\xi,\kappa_y,\kappa_z, \Omega) | =| \hat{\psi}(0,\kappa_y,\kappa_z, \Omega) |,
\end{equation}
whenever $\varepsilon \kappa_\perp / (2\hat{T}) \leq 1$ (i.e., propagating modes).

\subsection{\label{sec:The Lax series approach}The Lax series approach}

A Taylor expansion of Eq.~\eqref{eq:Illia solver:psi} in powers of $\kappa_\perp$ (around $\kappa_\perp=0$) and $\xi$ (around $\xi=0$), reveals that the general solution of the wave equation depends on powers of $\varepsilon$ \cite{Favier2017}. Motivated by this fact, in order to solve Eq.~\eqref{eq:psi:xiupsilonzeta:Fourier} one can express $\hat{\psi}$ in a series using $\varepsilon$ as expansion parameter \cite{Lax1975}. Because this perturbative approach is a rearrangement of a Taylor expansion, its convergence is thus guaranteed by Taylor's theorem for any $\varepsilon$ if high-order terms are calculated as explained below (i.e., satisfying Maxwell consistency, preserving the symmetry between electric and magnetic fields, and absence of evanescent modes). For linearly-polarized laser pulses, the transverse components (i.e., $\hat{\psi}_{E_y}$, $\hat{\psi}_{E_z}$, $\hat{\psi}_{B_y}$ and $\hat{\psi}_{B_z}$, generically denoted as $\hat{\psi}_\perp$) expand in even powers of $\varepsilon$ \cite{Davis1979}:
\begin{equation}\label{eq:Lax:psi transverse}
\hat{\psi}_\perp (\xi,\kappa_y,\kappa_z,\Omega) = \sum_{j=0}^{\infty} \varepsilon^{2j} \hat{\psi}_\perp^{(2j)}(\xi,\kappa_y,\kappa_z,\Omega),
\end{equation}
whereas the longitudinal components (i.e., $\hat{\psi}_{E_x}$ and $\hat{\psi}_{B_x}$, generically denoted as $\hat{\psi}_\parallel$) expand in odd powers of $\varepsilon$:
\begin{equation}\label{eq:Lax:psi longitudinal}
\hat{\psi}_\parallel (\xi,\kappa_y,\kappa_z,\Omega) = \sum_{j=0}^{\infty} \varepsilon^{2j+1} \hat{\psi}_\parallel^{(2j+1)} (\xi,\kappa_y,\kappa_z,\Omega),
\end{equation}
where the functions $\hat{\psi}_\perp^{(2j)}$ and $\hat{\psi}_\parallel^{(2j+1)}$ have to be determined.

\subsection{\label{sec:Lax series: Splitting the wave equation}Lax series: Splitting the wave equation}

If we substitute Eqs.~\eqref{eq:Lax:psi transverse}~and~\eqref{eq:Lax:psi longitudinal} into Eq.~\eqref{eq:psi:xiupsilonzeta:Fourier}, the wave equation is split into recursive equations. By doing so, the series~\eqref{eq:Lax:psi transverse}~and~\eqref{eq:Lax:psi longitudinal} satisfying the split Eqs.~\eqref{eq:splittingwaveeq:psiperp j=0}-\eqref{eq:splittingwaveeq:psiparallel j>0}, respectively, will verify the wave equation~\eqref{eq:psi:xiupsilonzeta:Fourier} and hence will be completely equivalent to Eq.~\eqref{eq:Illia solver:psi}.

The lowest order ($j=0$) corresponds to the paraxial equation:
\begin{equation}\label{eq:splittingwaveeq:psiperp j=0}
\left( \frac{{\rm i}\kappa_\perp^2}{4\, \hat{T}}  +  \partial_\xi    \right) \hat{\psi}_\perp^{(0)} = 0,
\end{equation}
\begin{equation}\label{eq:splittingwaveeq:psiparallel j=0}
\left( \frac{{\rm i}\kappa_\perp^2}{4\, \hat{T}}  +  \partial_\xi    \right) \hat{\psi}_\parallel^{(1)} = 0,
\end{equation}
where $\hat{\psi}_\perp^{(0)} = C_{0,\, \perp}^{(0)} {\rm e}^{-{\rm i}\frac{\kappa_\perp^2}{4\, \hat{T}} \xi}$ and $\hat{\psi}_\parallel^{(1)} = C_{0,\,\parallel}^{(1)} {\rm e}^{-{\rm i}\frac{\kappa_\perp^2}{4\, \hat{T}} \xi}$ are, respectively, their solutions. The coefficients $C_{0,\, \perp}^{(0)}=C_{0,\, \perp}^{(0)}(\kappa_y,\kappa_z, \Omega)$ and $C_{0,\,\parallel}^{(1)}=C_{0,\,\parallel}^{(1)}(\kappa_y,\kappa_z, \Omega)$ do not depend on $\xi$ (see their expressions for Hermite-Gaussian and Laguerre-Gaussian beams in Appendix~\ref{sec:sup_mat_prxl}).

High-order corrections ($j > 0$) verify:
\begin{equation}\label{eq:splittingwaveeq:psiperp j>0}
\left( \frac{{\rm i}\kappa_\perp^2}{4\, \hat{T}}  +  \partial_\xi    \right) \hat{\psi}_\perp^{(2j)} = \frac{{\rm i}}{4\, \hat{T}} \partial_\xi^2 \hat{\psi}_\perp^{(2j-2)},
\end{equation}
\begin{equation}\label{eq:splittingwaveeq:psiparallel j>0}
\left( \frac{{\rm i}\kappa_\perp^2}{4\, \hat{T}}  +  \partial_\xi    \right) \hat{\psi}_\parallel^{(2j+1)} = \frac{{\rm i}}{4\, \hat{T}} \partial_\xi^2 \hat{\psi}_\parallel^{(2j-1)},
\end{equation}
with the paraxial differential operator in the left-hand side. We choose to express the solution to Eqs.~\eqref{eq:splittingwaveeq:psiperp j>0}~and~\eqref{eq:splittingwaveeq:psiparallel j>0} as the sum of a homogeneous solution $\hat{H}$ and a particular solution $\hat{P}$:
\begin{equation}\label{eq:splittingwaveeq:psiperp:H+P}
\hat{\psi}_\perp^{(2j)} = \hat{H}_\perp^{(2j)} + \hat{P}_\perp^{(2j)},
\end{equation}
\begin{equation}\label{eq:splittingwaveeq:psiparallel:H+P}
\hat{\psi}_\parallel^{(2j+1)} =  \hat{H}_\parallel^{(2j+1)} + \hat{P}_\parallel^{(2j+1)},
\end{equation}
where the homogeneous solutions are, respectively:
\begin{equation}\label{eq:splittingwaveeq:psiperp:H}
\hat{H}_\perp^{(2j)} = C_{0,\, \perp}^{(2j)} {\rm e}^{-{\rm i}\frac{\kappa_\perp^2}{4\, \hat{T}} \xi},
\end{equation}
\begin{equation}\label{eq:splittingwaveeq:psiparallel:H}
\hat{H}_\parallel^{(2j+1)} =C_{0,\,\parallel}^{(2j+1)}  {\rm e}^{-{\rm i}\frac{\kappa_\perp^2}{4\, \hat{T}} \xi},
\end{equation}
where the coefficients $C_{0,\, \perp}^{(2j)}=C_{0,\, \perp}^{(2j)}(\kappa_y,\kappa_z, \Omega)$ and $C_{0,\,\parallel}^{(2j+1)}=C_{0,\,\parallel}^{(2j+1)}(\kappa_y,\kappa_z, \Omega)$ do not depend on $\xi$. It is important to note that even though $\hat{H}_\perp^{(2j)}$ and $\hat{H}_\parallel^{(2j+1)}$ formally obey the paraxial equation, they are part of the nonparaxial high-order corrections.

The particular solutions can be written as:
\begin{equation}\label{eq:splittingwaveeq:psiperp:P(xi)}
\hat{P}_\perp^{(2j)} = {\cal P}_\perp^{(2j)}(\xi) {\rm e}^{-{\rm i}\frac{\kappa_\perp^2}{4\, \hat{T}} \xi},
\end{equation}
\begin{equation}\label{eq:splittingwaveeq:psiparallel:P(xi)}
\hat{P}_\parallel^{(2j+1)} = {\cal P}_\parallel^{(2j+1)}(\xi)   {\rm e}^{-{\rm i}\frac{\kappa_\perp^2}{4\, \hat{T}} \xi},
\end{equation}
where the coefficients $ {\cal P}_\perp^{(2j)}(\xi)$ and ${\cal P}_\parallel^{(2j+1)}(\xi) $ do depend on $\xi$. Since in the neighborhood of the focal plane the form $\hat{\psi}\sim {\rm e}^{-{\rm i}\frac{\kappa_\perp^2}{4\, \hat{T}} \xi}$ dominates, the particular solutions must vanish in that plane, i.e., ${\cal P}_\perp^{(2j)}(0) = {\cal P}_\parallel^{(2j+1)}(0) =0$. To evaluate them through a recursive procedure as shown below, they can be constructed as $j$-order polynomials in $\xi$:
\begin{equation}\label{eq:splittingwaveeq:psiperp:P}
{\cal P}_\perp^{(2j)}(\xi) = \sum_{k=1}^j C_{k, \, \perp}^{(2j)} \, \xi^k,
\end{equation}
\begin{equation}\label{eq:splittingwaveeq:psiparallel:P}
{\cal P}_\parallel^{(2j+1)}(\xi) = \sum_{k=1}^j C_{k, \, \parallel}^{(2j+1)} \, \xi^k,
\end{equation}
where the coefficients $C_{k, \, \perp}^{(2j)}$ and $C_{k, \, \parallel}^{(2j+1)}$ have to be determined.

From the point of view of Lax recursive equations, homogeneous solutions $\hat{H}$ are simply arbitrary integration constants and hence Eqs.~\eqref{eq:splittingwaveeq:psiperp j>0}~and~\eqref{eq:splittingwaveeq:psiparallel j>0} do not suffice to determine them. These homogeneous solutions {\it must} be determined from the Maxwell equations by respecting the existing symmetry between the electric and magnetic fields (see Sec.~\ref{sec:Lax series: Splitting Maxwell equations}). We demonstrate in this paper that such Maxwell-consistent and symmetry-preserving calculation of the high-order corrections ensures that the overall laser energy through transverse planes is not modified by the Lax series terms of order $j>0$ (see Sec.~\ref{sec:Maxwell-consistent Lax series: properties}). This is a fundamental difference with respect to previous works, where, for example, in order to determine the high-order corrections, some authors had considered {\it ad hoc} assumptions such that they are zero at the beam focal point \cite{Couture1981, Takenaka1985}, they follow the structure of a spherical wave emanating from the beam focal point \cite{Salamin2007} or they must match some known nonparaxial solutions \cite{Favier2017}. Indeed, in the particular solutions proposed by most of these works dealing with Hermite-Gaussian and Laguerre-Gaussian paraxial families, spurious homogeneous solutions are found when a Gram-Schmidt orthogonalization process is applied in the focal plane \cite{Gram1883,Schmidt1907}. These spurious modes make the total power through transverse planes increase with $\varepsilon$ \cite{Salamin2007}, which is not physical.

\begin{figure*}
\includegraphics[width=\textwidth]{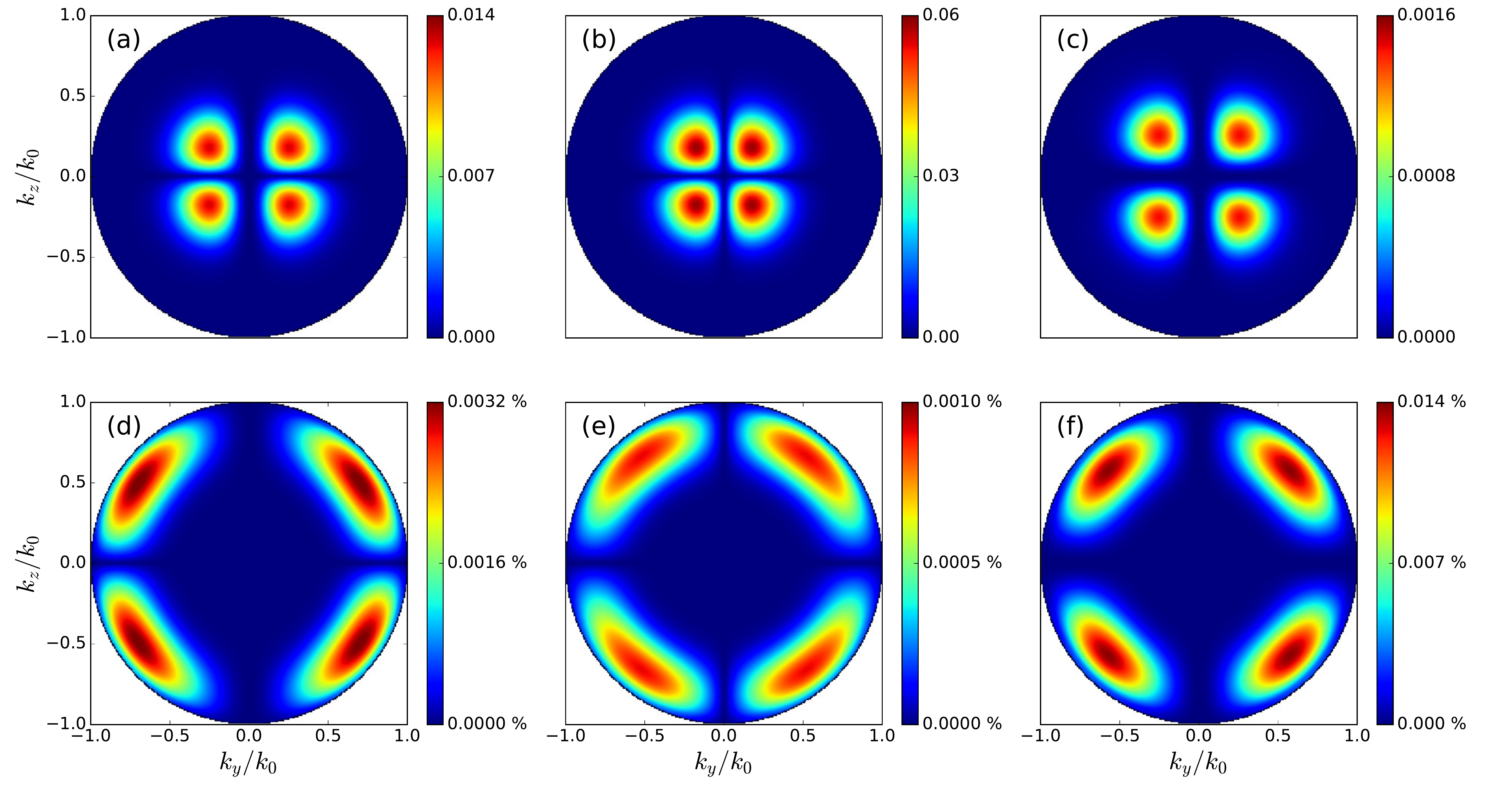}
\caption{\label{fig:HG11_errormap}Lax-series-based analytical solution $\hat{\psi}$ constructed from the $(1,1)$-order Hermite-Gaussian mode, truncated at order $j=5$. We consider $\lambda_0 = 800$~nm and $\varepsilon= 0.25$. The results are plotted in the transverse plane placed at $\xi = 1$. We show the spatial envelopes for (a) $E_x$, (b) $E_y$, and (c) $E_z$. The corresponding local relative errors, given by Eq.~\eqref{eq:example:def error local}, are shown in (d), (e) and (f), respectively.}
\end{figure*}

When substituting Eq.~\eqref{eq:splittingwaveeq:psiperp:H+P} into Eq.~\eqref{eq:splittingwaveeq:psiperp j>0}, and Eq.~\eqref{eq:splittingwaveeq:psiparallel:H+P} into Eq.~\eqref{eq:splittingwaveeq:psiparallel j>0}, the following recursion relations are obtained for the coefficients of the particular solutions for all $1\leq k \leq j$ and $j>0$:
\begin{equation}\label{eq:splittingwaveeq:psiperp:P recursion}
\begin{split}
C_{k,\,\perp}^{(2j)}  = &-\frac{{\rm i}\kappa_\perp^4}{64\, \hat{T}^3}\frac{C_{k-1,\,\perp}^{(2j-2)}}{k}+ \frac{\kappa_\perp^2}{8\, \hat{T}^2} C_{k,\,\perp}^{(2j-2)}  \\
& + \frac{\rm i}{4\, \hat{T}} (k+1) C_{k+1,\,\perp}^{(2j-2)},
\end{split}
\end{equation}
\begin{equation}\label{eq:splittingwaveeq:psiparallel:P recursion}
\begin{split}
C_{k,\,\parallel}^{(2j+1)}  = &-\frac{{\rm i}\kappa_\perp^4}{64\, \hat{T}^3}\frac{C_{k-1,\,\parallel}^{(2j-1)}}{k}+ \frac{\kappa_\perp^2}{8\, \hat{T}^2} C_{k,\,\parallel}^{(2j-1)}  \\
& + \frac{\rm i}{4\, \hat{T}} (k+1) C_{k+1,\,\parallel}^{(2j-1)},
\end{split}
\end{equation}
where, by notation convention, $ C_{k,\,\perp}^{(2j-2)} = C_{k,\,\parallel}^{(2j-1)} = 0$ if $k=j$ and $ C_{k+1,\,\perp}^{(2j-2)} = C_{k+1,\,\parallel}^{(2j-1)} = 0$ if $k \geq j-1$.

It is important to note that the above recursive relations involve the coefficients $C_{0,\,\perp}^{(2j-2)}$ and $C_{0,\,\parallel}^{(2j-1)}$ of the homogeneous solution, which will be determined from the Maxwell equations in the following subsection~\ref{sec:Lax series: Splitting Maxwell equations}.

\subsection{\label{sec:Lax series: Splitting Maxwell equations}Lax series: Splitting Maxwell equations}

We split Maxwell equations by substituting the Lax expansions~\eqref{eq:Lax:psi transverse}~and~\eqref{eq:Lax:psi longitudinal}, together with the Ans{\"a}tze~(\ref{eq:Ansatz:E})~and~(\ref{eq:Ansatz:B}), into Eqs.~\eqref{eq:divE=0}-\eqref{eq:Ampere}.

The envelopes of all the electromagnetic components at paraxial order ($j=0$) must verify simultaneously the following overdetermined system of equations:
\begin{equation}\label{eq:splitMaxwell:eq1 j=0}
\hat{T}\,\hat{\psi}_{E_x}^{(1)} + \frac{\kappa_y}{2} \hat{\psi}_{E_y}^{(0)} + \frac{\kappa_z}{2} \hat{\psi}_{E_z}^{(0)} = 0,
\end{equation}
\begin{equation}\label{eq:splitMaxwell:eq2 j=0}
\hat{T}\,\hat{\psi}_{B_x}^{(1)} + \frac{\kappa_y}{2} \hat{\psi}_{B_y}^{(0)} + \frac{\kappa_z}{2} \hat{\psi}_{B_z}^{(0)} = 0,
\end{equation}
\begin{equation}\label{eq:splitMaxwell:eq3 j=0}
\hat{T}\,\hat{\psi}_{B_x}^{(1)} - \frac{\kappa_y}{2} \hat{\psi}_{E_z}^{(0)} + \frac{\kappa_z}{2} \hat{\psi}_{E_y}^{(0)} = 0,
\end{equation}
\begin{equation}\label{eq:splitMaxwell:eq4 j=0}
\hat{\psi}_{B_y}^{(0)} + \hat{\psi}_{E_z}^{(0)} = 0,
\end{equation}
\begin{equation}\label{eq:splitMaxwell:eq5 j=0}
\hat{\psi}_{B_z}^{(0)} - \hat{\psi}_{E_y}^{(0)} = 0,
\end{equation}
\begin{equation}\label{eq:splitMaxwell:eq6 j=0}
\hat{T}\,\hat{\psi}_{E_x}^{(1)} + \frac{\kappa_y}{2} \hat{\psi}_{B_z}^{(0)} - \frac{\kappa_z}{2} \hat{\psi}_{B_y}^{(0)} = 0,
\end{equation}
which has a unique solution whatever two components are prescribed \cite{Thiele2016}. In this paper, without loss of generality, we choose the paraxial-order electric field polarized along $y$ axis (note that the solution for any other polarization angle can be obtained by applying a rotation transformation):
\begin{equation}\label{eq:splitMaxwell:Ey0}
\hat{\psi}_{E_y}^{(0)} = C \, {\rm e}^{-{\rm i}\frac{\kappa_\perp^2}{4\,\hat{T}} \xi},
\end{equation}
\begin{equation}\label{eq:splitMaxwell:Ez0}
\hat{\psi}_{E_z}^{(0)} = 0,
\end{equation}
where $C(\kappa_y,\kappa_z,\Omega)$ is a coefficient not depending on $\xi$. The rest of the components are then calculated from the system~\eqref{eq:splitMaxwell:eq1 j=0}-\eqref{eq:splitMaxwell:eq6 j=0}:
\begin{equation}\label{eq:splitMaxwell:Ex0}
\hat{\psi}_{E_x}^{(1)} =  -\frac{\kappa_y}{2\,\hat{T}}  \, C \, {\rm e}^{-{\rm i}\frac{\kappa_\perp^2}{4\,\hat{T}} \xi},
\end{equation}
\begin{equation}\label{eq:splitMaxwell:Bx0}
\hat{\psi}_{B_x}^{(1)} = -\frac{\kappa_z}{2\,\hat{T}}   \, C \, {\rm e}^{-{\rm i}\frac{\kappa_\perp^2}{4\,\hat{T}} \xi},
\end{equation}
\begin{equation}\label{eq:splitMaxwell:By0}
\hat{\psi}_{B_y}^{(0)} =  0,
\end{equation}
\begin{equation}\label{eq:splitMaxwell:Bz0}
\hat{\psi}_{B_z}^{(0)} = C \, {\rm e}^{-{\rm i}\frac{\kappa_\perp^2}{4\,\hat{T}} \xi}.
\end{equation}

\begin{figure*}
\includegraphics[width=\textwidth]{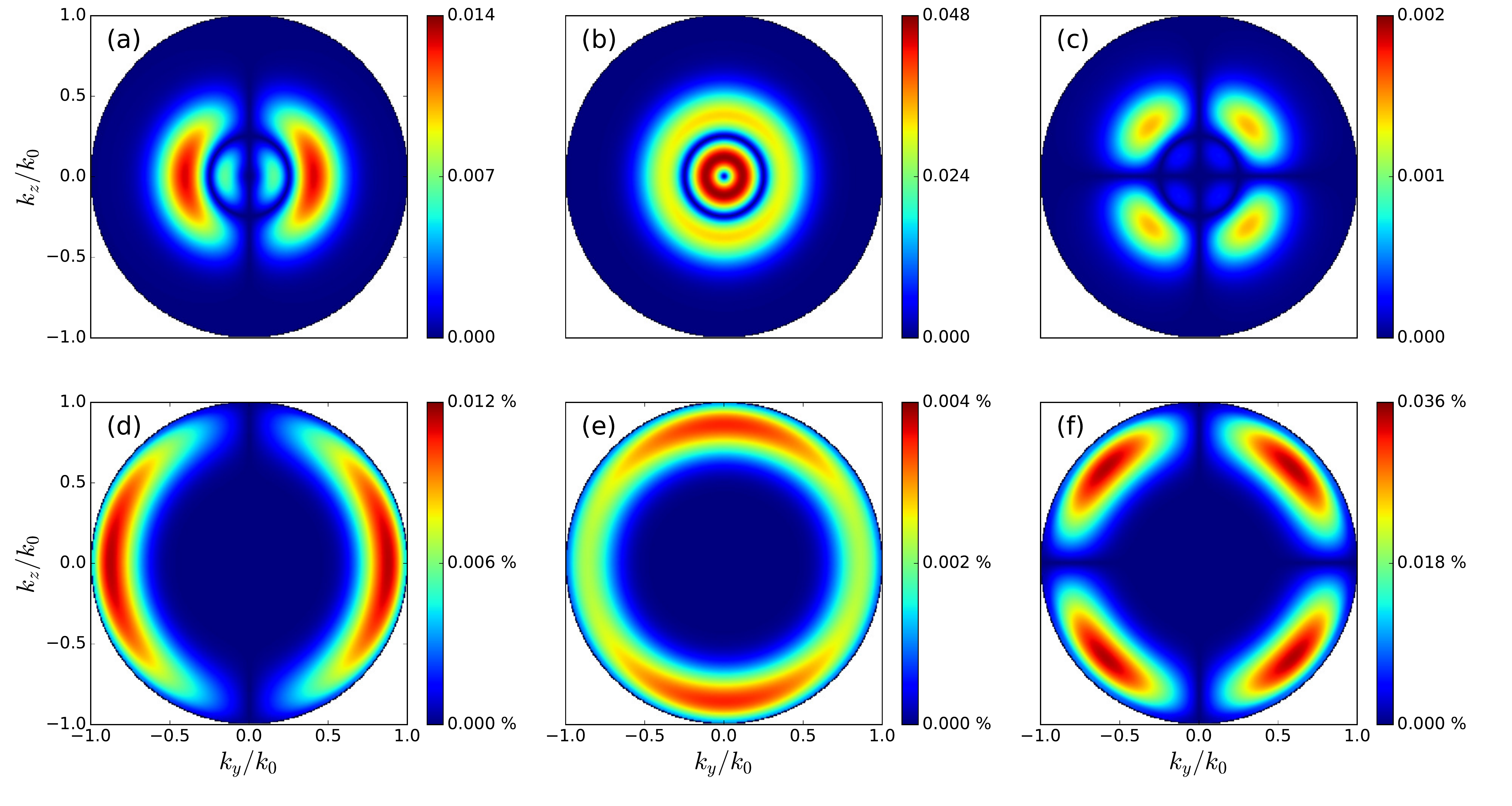}
  \caption{\label{fig:LG11_errormap}Lax-series-based analytical solution $\hat{\psi}$ constructed from the $(1,1)$-order Laguerre-Gaussian mode, truncated at order $j=5$. We consider $\lambda_0 = 800$~nm and $\varepsilon= 0.25$. The results are plotted in the transverse plane placed at $\xi = 1$. We show the spatial envelopes for (a) $E_x$, (b) $E_y$, and (c) $E_z$. The corresponding local relative errors, given by Eq.~\eqref{eq:example:def error local}, are shown in (d), (e) and (f), respectively.}
\end{figure*}

Similarly, the envelopes at high orders ($ j>0 $) must verify simultaneously the following overdetermined system of recursive equations:
\begin{equation}\label{eq:splitMaxwell:eq1 j>0}
2{\rm i}\,\hat{T}\,\hat{\psi}_{E_x}^{(2j+1)}    + {\rm i}\kappa_y \hat{\psi}_{E_y}^{(2j)} + {\rm i}{\kappa_z} \hat{\psi}_{E_z}^{(2j)} = - \partial_{\xi}  \hat{\psi}_{E_x}^{(2j-1)},
\end{equation}
\begin{equation}\label{eq:splitMaxwell:eq2 j>0}
2{\rm i}\,\hat{T}\,\hat{\psi}_{B_x}^{(2j+1)}  + {\rm i}\kappa_y \hat{\psi}_{B_y}^{(2j)} + {\rm i}{\kappa_z} \hat{\psi}_{B_z}^{(2j)} = -  \partial_{\xi}  \hat{\psi}_{B_x}^{(2j-1)} ,
\end{equation}
\begin{equation}\label{eq:splitMaxwell:eq3 j>0}
\hat{T}\,\hat{\psi}_{B_x}^{(2j+1)} - \frac{\kappa_y}{2} \hat{\psi}_{E_z}^{(2j)} + \frac{\kappa_z}{2} \hat{\psi}_{E_y}^{(2j)} = 0,
\end{equation}
\begin{equation}\label{eq:splitMaxwell:eq4 j>0}
\hat{\psi}_{B_y}^{(2j)} + \hat{\psi}_{E_z}^{(2j)} = \frac{\kappa_z}{2\,\hat{T}}  \hat{\psi}_{E_x}^{(2j-1)} + \frac{\rm i}{2\,\hat{T}} \partial_{\xi} \hat{\psi}_{E_z}^{(2j-2)}  ,
\end{equation}
\begin{equation}\label{eq:splitMaxwell:eq5 j>0}
\hat{\psi}_{B_z}^{(2j)} - \hat{\psi}_{E_y}^{(2j)} =- \frac{\kappa_y}{2\,\hat{T}}  \hat{\psi}_{E_x}^{(2j-1)} - \frac{\rm i}{2\,\hat{T}} \partial_{\xi} \hat{\psi}_{E_y}^{(2j-2)}  ,
\end{equation}
\begin{equation}\label{eq:splitMaxwell:eq6 j>0}
\hat{T}\,\hat{\psi}_{E_x}^{(2j+1)} + \frac{\kappa_y}{2} \hat{\psi}_{B_z}^{(2j)} - \frac{\kappa_z}{2} \hat{\psi}_{B_y}^{(2j)} = 0,
\end{equation}
\begin{equation}\label{eq:splitMaxwell:eq7 j>0}
\hat{\psi}_{B_z}^{(2j)} - \hat{\psi}_{E_y}^{(2j)} = \frac{\kappa_z}{2\,\hat{T}}  \hat{\psi}_{B_x}^{(2j-1)} + \frac{\rm i}{2\,\hat{T}} \partial_{\xi} \hat{\psi}_{B_z}^{(2j-2)}  ,
\end{equation}
\begin{equation}\label{eq:splitMaxwell:eq8 j>0}
\hat{\psi}_{B_y}^{(2j)} + \hat{\psi}_{E_z}^{(2j)} = \frac{\kappa_y}{2\,\hat{T}}  \hat{\psi}_{B_x}^{(2j-1)} + \frac{\rm i}{2\,\hat{T}} \partial_{\xi} \hat{\psi}_{B_y}^{(2j-2)}  ,
\end{equation}
which allows us to calculate the homogeneous parts in Eqs.~\eqref{eq:splittingwaveeq:psiperp:H+P}~and~\eqref{eq:splittingwaveeq:psiparallel:H+P}. Note that the particular solutions calculated in Sec.~\ref{sec:Lax series: Splitting the wave equation} satisfy all equations~\eqref{eq:splitMaxwell:eq1 j>0}-\eqref{eq:splitMaxwell:eq8 j>0}. In order to determine a unique homogeneous solution, we have to account for the symmetry existing between the electric and magnetic fields. For forward-propagating linearly-polarized pulses, by observing Eqs.~\eqref{eq:splitMaxwell:eq4 j>0}~and~\eqref{eq:splitMaxwell:eq8 j>0} and Eqs.~\eqref{eq:splitMaxwell:eq5 j>0}~and~\eqref{eq:splitMaxwell:eq7 j>0}, we require that:
\begin{equation}\label{eq:splitMaxwell:symmetry HEz HBy}
\hat{H}^{(2j)}_{B_y} - \hat{H}^{(2j)}_{E_z} =0,
\end{equation}
\begin{equation}\label{eq:splitMaxwell:symmetry HEy HBz}
\hat{H}^{(2j)}_{B_z} + \hat{H}^{(2j)}_{E_y} =0,
\end{equation}
which indeed is the opposite situation to the paraxial order (compare to Eqs.~\eqref{eq:splitMaxwell:eq4 j=0}~and~\eqref{eq:splitMaxwell:eq5 j=0}). {\it A posteriori}, we will demonstrate in Sec.~\ref{sec:Maxwell-consistent Lax series: properties} that this symmetry condition prevents high-order corrections from modifying the total laser energy.

After some manipulations, taking into account that we have prescribed the transverse electric field as in Eqs.~\eqref{eq:splitMaxwell:Ey0}~and~\eqref{eq:splitMaxwell:Ez0}, we get the following homogeneous solution for orders $j>0$:
\begin{equation}\label{eq:splitMaxwell:HEx j>0}
C_{0, \,E_x}^{(2j+1)}  =  \frac{\kappa_\perp^2}{16\,\hat{T}^2} C_{0, \,E_x}^{(2j-1)}  +\frac{\rm i}{4\,\hat{T}} C_{1, \,E_x}^{(2j-1)},
\end{equation}
\begin{equation}\label{eq:splitMaxwell:HEy j>0}
\begin{split}
C_{0,\,E_y}^{(2j)} & = \frac{\kappa_z^2}{8\,\hat{T}^2}C_{0,\,E_y}^{(2j-2)} -\frac{\kappa_\perp^2}{16\,\hat{T}^2}  C_{0,\,B_z}^{(2j-2)} \\
& - \frac{\kappa_y\kappa_z}{8\,\hat{T}^2} C_{0,\,E_z}^{(2j-2)} - \frac{\rm i}{4\,\hat{T}} C_{1,\,B_z}^{(2j-2)},
\end{split}
\end{equation}
\begin{equation}\label{eq:splitMaxwell:HEz j>0}
\begin{split}
C_{0,\,E_z}^{(2j)} & = \frac{\kappa_y^2}{8\,\hat{T}^2}C_{0,\,E_z}^{(2j-2)} +\frac{\kappa_\perp^2}{16\,\hat{T}^2}  C_{0,\,B_y}^{(2j-2)} \\
& - \frac{\kappa_y\kappa_z}{8\,\hat{T}^2} C_{0,\,E_y}^{(2j-2)} + \frac{\rm i}{4\,\hat{T}} C_{1,\,B_y}^{(2j-2)},
\end{split}
\end{equation}
\begin{equation}\label{eq:splitMaxwell:HBx j>0}
C_{0,\,B_x}^{(2j+1)} = \frac{\kappa_\perp^2}{16\,\hat{T}^2} C_{0,\,B_x}^{(2j-1)}  +\frac{\rm i}{4\,\hat{T}}C_{1,\,B_x}^{(2j-1)},
\end{equation}
\begin{equation}\label{eq:splitMaxwell:HBy j>0}
C_{0,\,B_y}^{(2j+1)} = C_{0,\,E_z}^{(2j)},
\end{equation}
\begin{equation}\label{eq:splitMaxwell:HBz j>0}
C_{0,\,B_z}^{(2j+1)} = - C_{0,\,E_y}^{(2j)},
\end{equation}
where $C_{1,\,E_x}^{(1)}=C_{1,\,B_x}^{(1)}=C_{1,\,B_y}^{(0)}=C_{1,\,B_z}^{(0)}=0$ by notation convention.

In conclusion, by setting $C$ in Eqs.~\eqref{eq:splitMaxwell:Ey0}~and~\eqref{eq:splitMaxwell:Ez0} the nonparaxial solution can be calculated in the whole space thanks to the recursive formulae Eqs.~\eqref{eq:splittingwaveeq:psiperp:P recursion}-\eqref{eq:splittingwaveeq:psiparallel:P recursion} and Eqs.~\eqref{eq:splitMaxwell:HEx j>0}-\eqref{eq:splitMaxwell:HBz j>0}. By way of example, the correction at order $j=1$ reads:
\begin{equation}\label{eq:splitMaxwell:Ex j=1}
\hat{\psi}_{E_x}^{(3)} =  \left[\frac{\kappa_\perp^2}{16\,\hat{T}^2}-\frac{{\rm i}\kappa_\perp^4}{64\,\hat{T}^3}\xi  \right] \hat{\psi}_{E_x}^{(1)},
\end{equation}
\begin{equation}\label{eq:splitMaxwell:Ey j=1}
\hat{\psi}_{E_y}^{(2)} = \left[\frac{\kappa_z^2-\kappa_y^2}{16\,\hat{T}^2}-\frac{{\rm i}\kappa_\perp^4}{64\,\hat{T}^3}\xi  \right] \hat{\psi}_{E_y}^{(0)} - \frac{\kappa_y\kappa_z}{8\,\hat{T}^2}\hat{\psi}_{E_z}^{(0)},
\end{equation}
\begin{equation}\label{eq:splitMaxwell:Ez j=1}
\hat{\psi}_{E_z}^{(2)} = \left[\frac{\kappa_y^2-\kappa_z^2}{16\,\hat{T}^2}-\frac{{\rm i}\kappa_\perp^4}{64\,\hat{T}^3}\xi  \right] \hat{\psi}_{E_z}^{(0)} - \frac{\kappa_y\kappa_z}{8\,\hat{T}^2}\hat{\psi}_{E_y}^{(0)},
\end{equation}
\begin{equation}\label{eq:splitMaxwell:Bx j=1}
\hat{\psi}_{B_x}^{(3)} =  \left[\frac{\kappa_\perp^2}{16\,\hat{T}^2}-\frac{{\rm i}\kappa_\perp^4}{64\,\hat{T}^3}\xi  \right] \hat{\psi}_{B_x}^{(1)},
\end{equation}
\begin{equation}\label{eq:splitMaxwell:By j=1}
\hat{\psi}_{B_y}^{(2)} = \left[\frac{\kappa_y^2-\kappa_z^2}{16\,\hat{T}^2}+\frac{{\rm i}\kappa_\perp^4}{64\,\hat{T}^3}\xi  \right] \hat{\psi}_{E_z}^{(0)} - \frac{\kappa_y\kappa_z}{8\,\hat{T}^2}\hat{\psi}_{E_y}^{(0)},
\end{equation}
\begin{equation}\label{eq:splitMaxwell:Bz j=1}
\hat{\psi}_{B_z}^{(2)} = \left[-\frac{\kappa_z^2-\kappa_y^2}{16\,\hat{T}^2}-\frac{{\rm i}\kappa_\perp^4}{64\,\hat{T}^3}\xi  \right] \hat{\psi}_{E_y}^{(0)} + \frac{\kappa_y\kappa_z}{8\,\hat{T}^2}\hat{\psi}_{E_z}^{(0)}.
\end{equation}

\subsection{Example: Monochromatic Hermite-Gaussian and Laguerre-Gaussian beams}

\begin{figure}
\centering
\includegraphics[width=\linewidth]{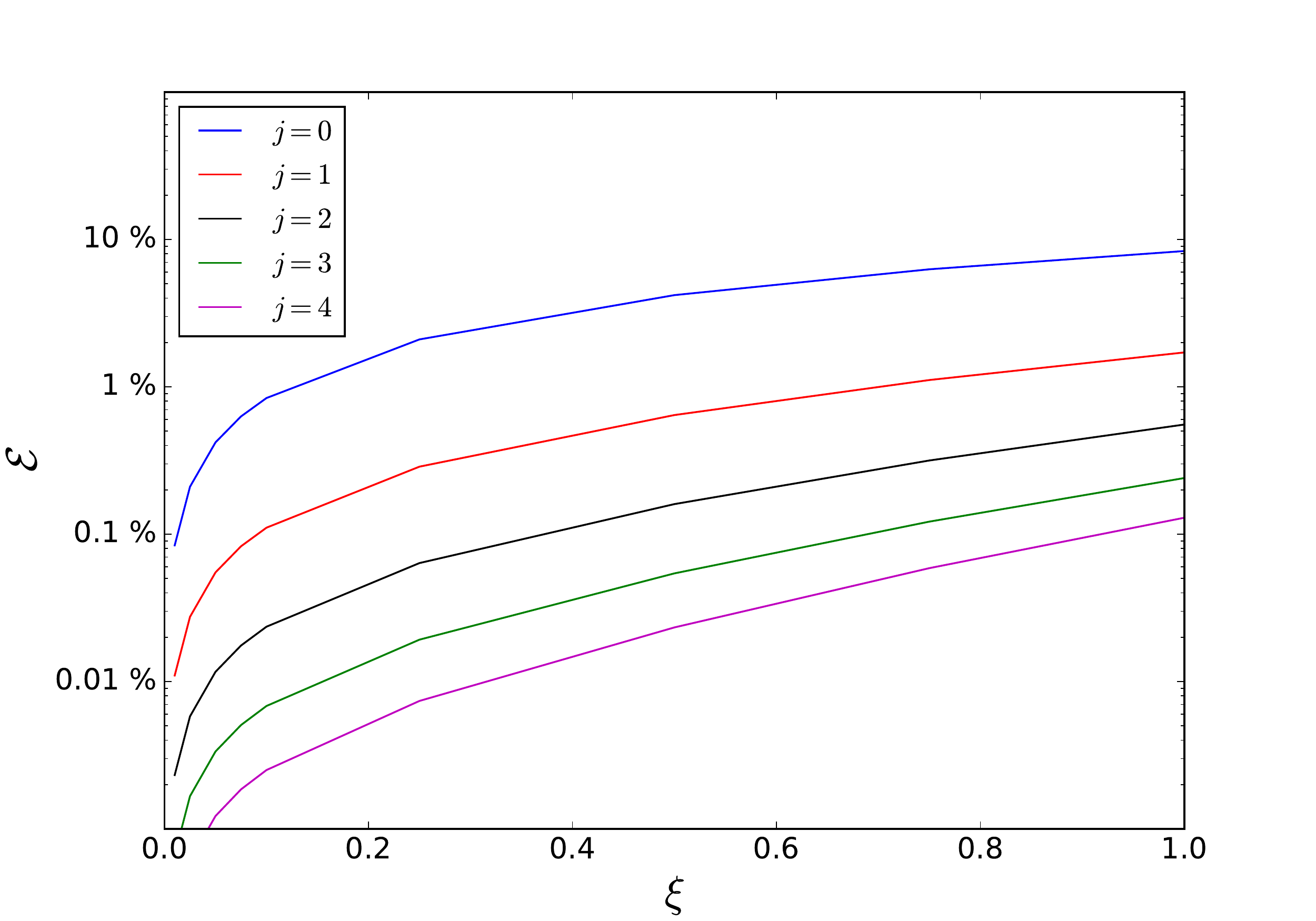}
\caption{\label{fig:error_xi_HG_11}Global relative error [Eq.~\eqref{eq:example:def error global}] between our analytical solution and the exact solution for $E_y$ as a function of the longitudinal coordinate at different truncation orders. The Lax series is built from the $(1,1)$-order Hermite-Gaussian mode, taking $\lambda_0 = 800$~nm and $\varepsilon= 0.25$.}
\end{figure}

We confront our Lax-series-based analytical solution to a numerical algorithm computing Maxwell-consistent solutions \cite{Thiele2016} (see Appendix~\ref{sec:sup_mat_Illia}). To do so, monochromatic beams are considered (i.e., $\hat{T}\rightarrow 1$) and the coefficient $C(\kappa_y,\kappa_z,\Omega)$ in Eqs.~\eqref{eq:splitMaxwell:Ey0}-\eqref{eq:splitMaxwell:Bz0} shall refer here to either a Hermite-Gaussian beam (see Eq.~\eqref{eq:sup_mat_prxl:HERMITE:C FFT HGnm}) or a Laguerre-Gaussian beam (see Eq.~\eqref{eq:sup_mat_prxl:LAGUERRE:C FFT LGpl}). Since the Lax series originates from a Taylor expansion around the beam focus, the best way to proceed is to prescribe our analytical solution in the focal plane, truncated at different orders $j$, and subsequently measure, for all electromagnetic components, the error between the solution of the exact solver ($\hat{\psi}^{\rm solver}$) and our analytical solution ($\hat{\psi}$) in different transverse planes. We compute errors using the standard Frobenius norm. The local relative error in a transverse plane is quantified as:
\begin{equation}\label{eq:example:def error local}
e = e(\xi,\kappa_y,\kappa_z) =  \frac{k_0\,\left| \hat{\psi}^{\rm solver} -  \hat{\psi} \right| }{\sqrt{   \displaystyle \iint_{k_\perp^2 \leq k_0^2} \left|  \hat{\psi}^{\rm solver} \right|^2  \,dk_y\,dk_z}      },
\end{equation}
and the global relative error in the same plane is:
\begin{equation}\label{eq:example:def error global}
{\cal E} = {\cal E}(\xi) = \sqrt{    \frac{\displaystyle \iint_{k_\perp^2 \leq k_0^2} \left| \hat{\psi}^{\rm solver} -  \hat{\psi} \right|^2  \,dk_y\,dk_z }{\displaystyle \iint_{k_\perp^2 \leq k_0^2} \left|  \hat{\psi}^{\rm solver} \right|^2  \,dk_y\,dk_z}      }.
\end{equation}

Figures~\ref{fig:HG11_errormap}~and~\ref{fig:LG11_errormap} show the analytical solution built from a $(1,1)$-order Hermite-Gaussian and $(1,1)$-order Laguerre-Gaussian modes, respectively, in the transverse plane placed at $\xi = 1$. We take $\lambda_0 = 800$~nm and a moderate $\varepsilon=0.25$ (for which the evanescent power is negligible). The highest local relative error (see Eq.~\eqref{eq:example:def error local}) appear in a ring (i.e., high values of transverse wavenumbers). When increasing the truncation order of the Lax series, this ring becomes narrower and the errors reduce in absolute value (not shown). This confirms numerically the convergence in the propagating region $k_\perp\leq k_0$ of our Lax-series-based solution seen as a Taylor expansion in $\kappa_y$ and $\kappa_z$. Figure~\ref{fig:error_xi_HG_11} shows that the global relative error diminishes too in all transverse planes when increasing the truncation order. This also confirms numerically the convergence in the propagating region of our Lax-series-based solution seen as a Taylor expansion in $\xi$. In the following subsection~\ref{sec:convergence} we shall demonstrate that our solution converges by giving the limit of the series for the six electromagnetic field components.

\subsection{\label{sec:convergence}Convergence of the solution}

The Ans{\"a}tze \eqref{eq:Ansatz:E} and \eqref{eq:Ansatz:B} are substituted into the Maxwell equations~\eqref{eq:divE=0}-\eqref{eq:Ampere}. In the transverse-spatial and temporal Fourier space, the resulting equations read:
\begin{equation}\label{eq:appendix:proofenergy:Maxwelleq:1}
{\rm i}\,\hat{T} \,\hat{\psi}_{E_x} + \frac{{\rm i}\,\varepsilon\,\kappa_y}{2}\,\hat{\psi}_{E_y} + \frac{{\rm i}\,\varepsilon\,\kappa_z}{2}\,\hat{\psi}_{E_z} = -\frac{\varepsilon^2}{2}\,\partial_\xi\hat{\psi}_{E_x}, 
\end{equation}
\begin{equation}\label{eq:appendix:proofenergy:Maxwelleq:2}
{\rm i}\,\hat{T} \,\hat{\psi}_{B_x} + \frac{{\rm i}\,\varepsilon\,\kappa_y}{2}\,\hat{\psi}_{B_y} + \frac{{\rm i}\,\varepsilon\,\kappa_z}{2}\,\hat{\psi}_{B_z} = -\frac{\varepsilon^2}{2}\,\partial_\xi\hat{\psi}_{B_x}, 
\end{equation}
\begin{equation}\label{eq:appendix:proofenergy:Maxwelleq:3}
{\rm i}\,\hat{T} \,\hat{\psi}_{B_x} - \frac{{\rm i}\,\varepsilon\,\kappa_y}{2}\,\hat{\psi}_{E_z} + \frac{{\rm i}\,\varepsilon\,\kappa_z}{2}\,\hat{\psi}_{E_y} = 0, 
\end{equation}
\begin{equation}\label{eq:appendix:proofenergy:Maxwelleq:4}
{\rm i}\,\hat{T} \,\hat{\psi}_{E_x} + \frac{{\rm i}\,\varepsilon\,\kappa_y}{2}\,\hat{\psi}_{B_z} - \frac{{\rm i}\,\varepsilon\,\kappa_z}{2}\,\hat{\psi}_{B_y} = 0, 
\end{equation}
\begin{equation}\label{eq:appendix:proofenergy:Maxwelleq:5}
{\rm i}\,\hat{T} \, \left(  \hat{\psi}_{B_y} + \hat{\psi}_{E_z} \right) = \frac{{\rm i}\,\varepsilon\,\kappa_z}{2}\,\hat{\psi}_{E_x}  -\frac{\varepsilon^2}{2}\,\partial_\xi\hat{\psi}_{E_z}, 
\end{equation}
\begin{equation}\label{eq:appendix:proofenergy:Maxwelleq:6}
{\rm i}\,\hat{T} \, \left(  \hat{\psi}_{B_y} + \hat{\psi}_{E_z} \right) = \frac{{\rm i}\,\varepsilon\,\kappa_y}{2}\,\hat{\psi}_{B_x}  -\frac{\varepsilon^2}{2}\,\partial_\xi\hat{\psi}_{B_y}, 
\end{equation}
\begin{equation}\label{eq:appendix:proofenergy:Maxwelleq:7}
{\rm i}\,\hat{T} \, \left(  \hat{\psi}_{B_z} - \hat{\psi}_{E_y} \right) = - \frac{{\rm i}\,\varepsilon\,\kappa_y}{2}\,\hat{\psi}_{E_x}  +\frac{\varepsilon^2}{2}\,\partial_\xi\hat{\psi}_{E_y}, 
\end{equation}
\begin{equation}\label{eq:appendix:proofenergy:Maxwelleq:8}
{\rm i}\,\hat{T} \, \left(  \hat{\psi}_{B_z} - \hat{\psi}_{E_y} \right) = \frac{{\rm i}\,\varepsilon\,\kappa_z}{2}\,\hat{\psi}_{B_x}  -\frac{\varepsilon^2}{2}\,\partial_\xi\hat{\psi}_{B_z}, 
\end{equation}
which, when they are split through Lax series~\eqref{eq:Lax:psi transverse}-\eqref{eq:Lax:psi longitudinal}, encompass Eqs.~\eqref{eq:splitMaxwell:eq1 j=0}-\eqref{eq:splitMaxwell:eq6 j=0} and Eqs.~\eqref{eq:splitMaxwell:eq1 j>0}-\eqref{eq:splitMaxwell:eq8 j>0}.

As explained in Sec.~\ref{sec:Lax series: Splitting the wave equation}, each envelope in Eqs.~\eqref{eq:appendix:proofenergy:Maxwelleq:1}-\eqref{eq:appendix:proofenergy:Maxwelleq:8} is assumed to be a forward-propagating solution of the wave equation~\eqref{eq:psi:xiupsilonzeta:Fourier}, which is given by Eq.~\eqref{eq:Illia solver:psi}. After some manipulations, the substitution of the form~\eqref{eq:Illia solver:psi}, whose boundary condition is placed at $ \xi=0$, into Eqs.~\eqref{eq:appendix:proofenergy:Maxwelleq:1}-\eqref{eq:appendix:proofenergy:Maxwelleq:8} yields:
\begin{equation}\label{eq:appendix:proofenergy:Maxwelleq+waveeq:1}
\frac{2}{\varepsilon}\, \hat{T} \,\mathscr{P}\,\hat{\psi}_{E_x}(0) + \kappa_y\,\hat{\psi}_{E_y}(0) + \kappa_z\,\hat{\psi}_{E_z}(0) = 0, 
\end{equation}
\begin{equation}\label{eq:appendix:proofenergy:Maxwelleq+waveeq:2}
\frac{2}{\varepsilon}\, \hat{T} \,\mathscr{P}\,\hat{\psi}_{B_x}(0) + \kappa_y\,\hat{\psi}_{B_y}(0) + \kappa_z\,\hat{\psi}_{B_z}(0) = 0, 
\end{equation}
\begin{equation}\label{eq:appendix:proofenergy:Maxwelleq+waveeq:3}
\frac{2}{\varepsilon}\, \hat{T} \,\hat{\psi}_{B_x}(0) - \kappa_y\,\hat{\psi}_{E_z}(0) + \kappa_z\,\hat{\psi}_{E_y}(0) = 0, 
\end{equation}
\begin{equation}\label{eq:appendix:proofenergy:Maxwelleq+waveeq:4}
\frac{2}{\varepsilon}\, \hat{T} \,\hat{\psi}_{E_x}(0) + \kappa_y\,\hat{\psi}_{B_z}(0) - \kappa_z\,\hat{\psi}_{B_y}(0) = 0, 
\end{equation}
\begin{equation}\label{eq:appendix:proofenergy:Maxwelleq+waveeq:5}
\begin{split}
&\frac{2}{\varepsilon}\, \hat{T} \, (1+\mathscr{P}) \,\left( \hat{\psi}_{B_y}(0) + \hat{\psi}_{E_z}(0) \right) = \\
& \kappa_y\,\hat{\psi}_{B_x}(0) + \kappa_z\,\hat{\psi}_{E_x}(0) , 
\end{split}
\end{equation}
\begin{equation}\label{eq:appendix:proofenergy:Maxwelleq+waveeq:6}
\begin{split}
&\frac{2}{\varepsilon}\, \hat{T} \, (1+\mathscr{P}) \,\left( \hat{\psi}_{B_z}(0) - \hat{\psi}_{E_y}(0) \right) = \\
& \kappa_z\,\hat{\psi}_{B_x}(0) - \kappa_y\,\hat{\psi}_{E_x}(0) , 
\end{split}
\end{equation}
where $\hat{\psi}(0)$ refers to the value of the corresponding envelope in the focal plane and the operator $\mathscr{P}$ is given by:
\begin{equation}
\mathscr{P} = \sqrt{1 - \frac{\varepsilon^2\,\kappa_\perp^2}{4\,\hat{T}^2}},
\end{equation}
where the argument of the square root must be nonnegative for forward-propagating waves (i.e., $\varepsilon \kappa_\perp /(2\hat{T})\leq 1$). Therefore $0 \leq \mathscr{P} \leq 1$, where the upper limit $\mathscr{P}\rightarrow 1$ represents the paraxial limit ($\varepsilon \rightarrow 0$).

Following Sec.~\ref{sec:Lax series: Splitting Maxwell equations}, one needs to impose the symmetry conditions~\eqref{eq:splitMaxwell:symmetry HEz HBy}~and~\eqref{eq:splitMaxwell:symmetry HEy HBz} in order to have a unique solution of Eqs.~\eqref{eq:appendix:proofenergy:Maxwelleq+waveeq:1}-\eqref{eq:appendix:proofenergy:Maxwelleq+waveeq:6}. The values of the envelopes of the transverse field components in the focal plane are thus:
\begin{equation}\label{eq:appendix:proofenergy:Ey xi=0}
\hat{\psi}_{E_y}(0) = C_{0,E_y}^{(0)} +\mathscr{H}_y,
\end{equation}
\begin{equation}\label{eq:appendix:proofenergy:Bz xi=0}
\hat{\psi}_{B_z}(0) = C_{0,E_y}^{(0)} -\mathscr{H}_y,
\end{equation}
\begin{equation}\label{eq:appendix:proofenergy:Ez xi=0}
\hat{\psi}_{E_z}(0) = C_{0,E_z}^{(0)} +\mathscr{H}_z,
\end{equation}
\begin{equation}\label{eq:appendix:proofenergy:By xi=0}
\hat{\psi}_{B_y}(0) = -C_{0,E_z}^{(0)} +\mathscr{H}_z,
\end{equation}
where the sum of the homogeneous parts of the high-order terms of the Lax series are:
\begin{equation}\label{eq:appendix:proofenergy:HEy}
\mathscr{H}_y = \sum_{j=1}^{\infty} \varepsilon^{2j}\, C_{0,E_y}^{(2j)},
\end{equation}
\begin{equation}\label{eq:appendix:proofenergy:HEy}
\mathscr{H}_z = \sum_{j=1}^{\infty} \varepsilon^{2j}\, C_{0,E_z}^{(2j)}.
\end{equation}

When inserting Eqs.~\eqref{eq:appendix:proofenergy:Ey xi=0}-\eqref{eq:appendix:proofenergy:By xi=0} into Eqs.~\eqref{eq:appendix:proofenergy:Maxwelleq+waveeq:1}-\eqref{eq:appendix:proofenergy:Maxwelleq+waveeq:6}, a unique solution is obtained in terms of $C_{0,E_y}^{(0)}$ and $C_{0,E_z}^{(0)}$:
\begin{equation}\label{eq:appendix:proofenergy:sol HEy}
\mathscr{H}_y =- \frac{\varepsilon^2\,(\kappa_y^2-\kappa_z^2) }{4\,\hat{T}^2(1+\mathscr{P})^2} \, C_{0,E_y}^{(0)} - \frac{\varepsilon^2\,\kappa_y\kappa_z }{2\,\hat{T}^2(1+\mathscr{P})^2}\, C_{0,E_z}^{(0)} ,
\end{equation}
\begin{equation}\label{eq:appendix:proofenergy:sol HEz}
\mathscr{H}_z = \frac{\varepsilon^2\,(\kappa_y^2-\kappa_z^2) }{4\,\hat{T}^2(1+\mathscr{P})^2} \, C_{0,E_z}^{(0)} - \frac{\varepsilon^2\,\kappa_y\kappa_z }{2\,\hat{T}^2(1+\mathscr{P})^2}\, C_{0,E_y}^{(0)} ,
\end{equation}
which yields:
\begin{equation}\label{eq:appendix:proofenergy:sol Ey xi=0}
\begin{split}
\hat{\psi}_{E_y}(0) = & \left[ 1 - \frac{\varepsilon^2\,(\kappa_y^2-\kappa_z^2) }{4\,\hat{T}^2(1+\mathscr{P})^2} \right] \, C_{0,E_y}^{(0)} \\
& - \frac{\varepsilon^2\,\kappa_y\kappa_z }{2\,\hat{T}^2(1+\mathscr{P})^2}  \, C_{0,E_z}^{(0)},
\end{split}
\end{equation}
\begin{equation}\label{eq:appendix:proofenergy:sol Bz xi=0}
\begin{split}
\hat{\psi}_{B_z}(0) = & \left[ 1 + \frac{\varepsilon^2\,(\kappa_y^2-\kappa_z^2) }{4\,\hat{T}^2(1+\mathscr{P})^2} \right] \, C_{0,E_y}^{(0)} \\
& + \frac{\varepsilon^2\,\kappa_y\kappa_z }{2\,\hat{T}^2(1+\mathscr{P})^2}  \, C_{0,E_z}^{(0)},
\end{split}
\end{equation}
\begin{equation}\label{eq:appendix:proofenergy:sol Ez xi=0}
\begin{split}
\hat{\psi}_{E_z}(0) = & \left[ 1 + \frac{\varepsilon^2\,(\kappa_y^2-\kappa_z^2) }{4\,\hat{T}^2(1+\mathscr{P})^2} \right] \, C_{0,E_z}^{(0)} \\
& - \frac{\varepsilon^2\,\kappa_y\kappa_z }{2\,\hat{T}^2(1+\mathscr{P})^2}  \, C_{0,E_y}^{(0)},
\end{split}
\end{equation}
\begin{equation}\label{eq:appendix:proofenergy:sol By xi=0}
\begin{split}
\hat{\psi}_{B_y}(0) = & \left[ -1 + \frac{\varepsilon^2\,(\kappa_y^2-\kappa_z^2) }{4\,\hat{T}^2(1+\mathscr{P})^2} \right] \, C_{0,E_z}^{(0)} \\
& - \frac{\varepsilon^2\,\kappa_y\kappa_z }{2\,\hat{T}^2(1+\mathscr{P})^2}  \, C_{0,E_y}^{(0)},
\end{split}
\end{equation}
\begin{equation}\label{eq:appendix:proofenergy:sol Ex xi=0}
\hat{\psi}_{E_x}(0) = - \frac{\varepsilon}{\hat{T}(1+\mathscr{P})} \left[ \kappa_y \, C_{0,E_y}^{(0)}  +   \kappa_z   C_{0,E_z}^{(0)} \right],
\end{equation}
\begin{equation}\label{eq:appendix:proofenergy:sol Bx xi=0}
\hat{\psi}_{B_x}(0) = - \frac{\varepsilon}{\hat{T}(1+\mathscr{P})} \left[ \kappa_z \, C_{0,E_y}^{(0)}  -   \kappa_y   C_{0,E_z}^{(0)} \right].
\end{equation}

The existence of the solutions~\eqref{eq:appendix:proofenergy:sol Ey xi=0}-\eqref{eq:appendix:proofenergy:sol Bx xi=0}, which result in {\it finite} values for $0\leq \mathscr{P} \leq 1$ (forward-propagating modes), implies that the Lax series obtained following our procedure is convergent, since these values actually represent the limit towards which our analytical solution converges in the focal plane.

Taking into account that paraxial-order terms follow Eq.~\eqref{eq:sup_mat_prxl:paraxial kykz} and solutions of the wave equation are governed by Eq.~\eqref{eq:Illia solver:psi}, from Eqs.~\eqref{eq:appendix:proofenergy:sol Ey xi=0}-\eqref{eq:appendix:proofenergy:sol Bx xi=0} one can express our solution as a function of the prescribed paraxial modes $\hat{\psi}_{E_y}^{(0)}$ and $\hat{\psi}_{E_z}^{(0)}$:
\begin{equation}\label{eq:appendix:proofenergy:sol Ey}
\begin{split}
\hat{\psi}_{E_y} = & \left[ 1 - \frac{\varepsilon^2\,(\kappa_y^2-\kappa_z^2) }{4\,\hat{T}^2(1+\mathscr{P})^2} \right] \, {\rm e}^{\frac{{\rm i}\, \hat{T} \, (1-\mathscr{P})}{\varepsilon^2}\xi}\, \hat{\psi}_{E_y}^{(0)} \\
& - \frac{\varepsilon^2\,\kappa_y\kappa_z }{2\,\hat{T}^2(1+\mathscr{P})^2}  \, {\rm e}^{\frac{{\rm i}\, \hat{T} \, (1-\mathscr{P})}{\varepsilon^2}\xi}\, \hat{\psi}_{E_z}^{(0)},
\end{split}
\end{equation}
\begin{equation}\label{eq:appendix:proofenergy:sol Bz}
\begin{split}
\hat{\psi}_{B_z} = & \left[ 1 + \frac{\varepsilon^2\,(\kappa_y^2-\kappa_z^2) }{4\,\hat{T}^2(1+\mathscr{P})^2} \right] \,{\rm e}^{\frac{{\rm i}\, \hat{T} \, (1-\mathscr{P})}{\varepsilon^2}\xi}\, \hat{\psi}_{E_y}^{(0)} \\
& + \frac{\varepsilon^2\,\kappa_y\kappa_z }{2\,\hat{T}^2(1+\mathscr{P})^2}  \,{\rm e}^{\frac{{\rm i}\, \hat{T} \, (1-\mathscr{P})}{\varepsilon^2}\xi}\, \hat{\psi}_{E_z}^{(0)},
\end{split}
\end{equation}
\begin{equation}\label{eq:appendix:proofenergy:sol Ez}
\begin{split}
\hat{\psi}_{E_z} = & \left[ 1 + \frac{\varepsilon^2\,(\kappa_y^2-\kappa_z^2) }{4\,\hat{T}^2(1+\mathscr{P})^2} \right] \, {\rm e}^{\frac{{\rm i}\, \hat{T} \, (1-\mathscr{P})}{\varepsilon^2}\xi}\,\hat{\psi}_{E_z}^{(0)} \\
& - \frac{\varepsilon^2\,\kappa_y\kappa_z }{2\,\hat{T}^2(1+\mathscr{P})^2}  \, {\rm e}^{\frac{{\rm i}\, \hat{T} \, (1-\mathscr{P})}{\varepsilon^2}\xi}\, \hat{\psi}_{E_y}^{(0)},
\end{split}
\end{equation}
\begin{equation}\label{eq:appendix:proofenergy:sol By}
\begin{split}
\hat{\psi}_{B_y} = & \left[ -1 + \frac{\varepsilon^2\,(\kappa_y^2-\kappa_z^2) }{4\,\hat{T}^2(1+\mathscr{P})^2} \right] \, {\rm e}^{\frac{{\rm i}\, \hat{T} \, (1-\mathscr{P})}{\varepsilon^2}\xi}\,\hat{\psi}_{E_z}^{(0)} \\
& - \frac{\varepsilon^2\,\kappa_y\kappa_z }{2\,\hat{T}^2(1+\mathscr{P})^2}  \, {\rm e}^{\frac{{\rm i}\, \hat{T} \, (1-\mathscr{P})}{\varepsilon^2}\xi}\, \hat{\psi}_{E_y}^{(0)},
\end{split}
\end{equation}
\begin{equation}\label{eq:appendix:proofenergy:sol Ex}
\hat{\psi}_{E_x} = - \frac{\varepsilon \, {\rm e}^{\frac{{\rm i}\, \hat{T} \, (1-\mathscr{P})}{\varepsilon^2}\xi}}{\hat{T}(1+\mathscr{P})} \left[ \kappa_y \, \hat{\psi}_{E_y}^{(0)}  +   \kappa_z   \hat{\psi}_{E_z}^{(0)} \right],
\end{equation}
\begin{equation}\label{eq:appendix:proofenergy:sol Bx}
\hat{\psi}_{B_x} = - \frac{\varepsilon \, {\rm e}^{\frac{{\rm i}\, \hat{T} \, (1-\mathscr{P})}{\varepsilon^2}\xi}}{\hat{T}(1+\mathscr{P})} \left[ \kappa_z \, \hat{\psi}_{E_y}^{(0)}  -   \kappa_y   \hat{\psi}_{E_z}^{(0)} \right].
\end{equation}

One can verify that a Taylor expansion in $\varepsilon$ of Eqs.~\eqref{eq:appendix:proofenergy:sol Ey}-\eqref{eq:appendix:proofenergy:sol Bx} yields the terms of our series presented in Secs.~\ref{sec:Lax series: Splitting the wave equation}~and~\ref{sec:Lax series: Splitting Maxwell equations}.

Inversely, assuming known a full forward-propagating solution of Maxwell equations, the underlying paraxial level from which that solution is constructed through the Lax series can be easily determined from Eqs.~\eqref{eq:appendix:proofenergy:sol Ey}~and~\eqref{eq:appendix:proofenergy:sol Ez}:
\begin{equation}\label{eq:appendix:proofenergy:sol Ey0}
\begin{split}
\hat{\psi}_{E_y}^{(0)} = & \left[ \frac{1+\mathscr{P}}{2} + \frac{\varepsilon^2\,\kappa_y^2 }{8\,\hat{T}^2\mathscr{P}} \right] \, {\rm e}^{-\frac{{\rm i}\, \hat{T} \, (1-\mathscr{P})}{\varepsilon^2}\xi}\,\hat{\psi}_{E_y} \\
& + \frac{\varepsilon^2\,\kappa_y\kappa_z }{8\,\hat{T}^2\mathscr{P}}  \, {\rm e}^{-\frac{{\rm i}\, \hat{T} \, (1-\mathscr{P})}{\varepsilon^2}\xi}\, \hat{\psi}_{E_z},
\end{split}
\end{equation}
\begin{equation}\label{eq:appendix:proofenergy:sol Ez0}
\begin{split}
\hat{\psi}_{E_z}^{(0)} = & \left[ \frac{1+\mathscr{P}}{2} + \frac{\varepsilon^2\,\kappa_z^2 }{8\,\hat{T}^2\mathscr{P}} \right] \, {\rm e}^{-\frac{{\rm i}\, \hat{T} \, (1-\mathscr{P})}{\varepsilon^2}\xi}\,\hat{\psi}_{E_z} \\
& + \frac{\varepsilon^2\,\kappa_y\kappa_z }{8\,\hat{T}^2\mathscr{P}}  \, {\rm e}^{-\frac{{\rm i}\, \hat{T} \, (1-\mathscr{P})}{\varepsilon^2}\xi}\, \hat{\psi}_{E_y},
\end{split}
\end{equation}
which needs the boundary condition $\hat{\psi} \rightarrow 0$ and $\hat{\psi}_\perp^{(0)}\rightarrow 0$ at $\kappa_\perp/\hat{T} = 2/\varepsilon$ (i.e., the separation between propagating and evanescent modes, given by $\mathscr{P}\rightarrow 0$).

Equations~\eqref{eq:appendix:proofenergy:sol Ey}-\eqref{eq:appendix:proofenergy:sol Bx} can be directly exploited to accurately inject tightly focused ultrashort laser pulses of arbitrary shape in space and time in Maxwell codes based on the Unidirectional Pulse Propagation Equation \cite{Kolesik2002,Kolesik2004}. Under the cost of computing an inverse Fourier transform in the transverse space and time \cite{Thiele2016}, these equations can also be used to prescribe the laser field under highly nonparaxial conditions on the boundaries of Finite-Difference-Time-Domain (FDTD) codes such as Particle-In-Cell (PIC) ones \cite{Birdsall1985,Hockney1988}. Since the spectrum is analytically known everywhere, the most efficient fashion of Fourier-backtransforming Eqs.~\eqref{eq:appendix:proofenergy:sol Ey}-\eqref{eq:appendix:proofenergy:sol Bx} is through Inverse Discrete Fourier Transforms (IDFT) based on quadrature formulae (see Sec.~\ref{sec:Asymptotic behavior far from focal plane}).

In the following subsection~\ref{sec:Maxwell-consistent Lax series: properties} we shall demonstrate that our solution conserves the energy.

\subsection{\label{sec:Maxwell-consistent Lax series: properties}Energy conservation}

The overall laser energy is calculated by integrating the longitudinal component of the Poynting vector ($\Pi_x$) over transverse coordinates and time (see Appendix~\ref{sec:sup_mat_innerprod}):
\begin{equation}\label{eq:appendix:proofenergy:energy:def}
U = \frac{D_0^2}{4\, \omega_0}  \iiint_{-\infty}^{+\infty} \Pi_x \, \, d\upsilon \, d\zeta \, d\tau,
\end{equation}
where $\Pi_x = c^2\varepsilon_0 (E_y \bar{B}_z - E_z \bar{B}_y)$. Taking into account the Ans{\"a}tze~\eqref{eq:Ansatz:E} and~\eqref{eq:Ansatz:B}, the normalized total energy expresses in terms of the inner product between envelopes defined by Eq.~\eqref{eq:sup_mat_innerprod:definition inner prod} as follows:
\begin{equation}\label{eq:appendix:proofenergy:energy:inner prod def}
\frac{4 \, \omega_0}{c\varepsilon_0E_0^2D_0^2}\,U = \langle \psi_{E_y} , \psi_{B_z} \rangle - \langle \psi_{E_z} , \psi_{B_y} \rangle.
\end{equation}

When substituting the Lax series~\eqref{eq:Lax:psi transverse} into Eq.~\eqref{eq:appendix:proofenergy:energy:inner prod def}, the overall energy expands in powers of $\varepsilon$ as follows:
\begin{equation}\label{eq:appendix:proofenergy:energy:Lax expansion}
\begin{split}
&\langle \psi_{E_y} , \psi_{B_z} \rangle - \langle \psi_{E_z} , \psi_{B_y} \rangle =\\
& \sum_{j=0}^{\infty}\varepsilon^{2j}\sum_{\alpha=0}^{j}  \langle \psi_{E_y}^{(2\alpha)} , \psi_{B_z}^{(2j-2\alpha)} \rangle  -\\
& \sum_{j=0}^{\infty}\varepsilon^{2j}\sum_{\alpha=0}^{j}  \langle \psi_{E_z}^{(2\alpha)} , \psi_{B_y}^{(2j-2\alpha)} \rangle.
\end{split}
\end{equation}

In order to demonstrate the energy conservation, we shall search for the least upper and lower bounds of the total energy. From Eq.~\eqref{eq:appendix:proofenergy:energy:Lax expansion} one easily deduces that the total energy is bounded from below by the paraxial-order energy:
\begin{equation}\label{eq:appendix:proofenergy:energy:lower bound}
\begin{split}
&\langle \psi_{E_y} , \psi_{B_z} \rangle - \langle \psi_{E_z} , \psi_{B_y} \rangle \geq \\
&\langle \psi_{E_y}^{(0)} , \psi_{B_z}^{(0)} \rangle - \langle \psi_{E_z}^{(0)} , \psi_{B_y}^{(0)} \rangle,
\end{split}
\end{equation}
since, for forward-propagating waves, no $\varepsilon$-order term in Eq.~\eqref{eq:appendix:proofenergy:energy:Lax expansion} can be negative. 

We shall seek the least upper bound for the total energy in the transverse-spatial and temporal Fourier space. 
Thanks to the Plancherel's theorem, the normalized overall energy~\eqref{eq:appendix:proofenergy:energy:inner prod def} can be calculated in the Fourier space as follows:
\begin{equation}\label{eq:appendix:proofenergy:energy:Plancherel}
\begin{split}
&\frac{4 \, \omega_0}{c\varepsilon_0E_0^2D_0^2}\,U =  \\
& 8\pi^3  \iiint_{\frac{\varepsilon \kappa_\perp}{2\hat{T}}\leq 1} \left( \hat{\psi}_{ E_y} \bar{\hat{\psi}}_{B_z} - \hat{\psi}_{E_z} \bar{\hat{\psi}}_{B_y}  \right) \, d\kappa_y \, d\kappa_z \, d\Omega.
\end{split}
\end{equation}

From the solutions~\eqref{eq:appendix:proofenergy:sol Ey}-\eqref{eq:appendix:proofenergy:sol By} the integrand in Eq.~\eqref{eq:appendix:proofenergy:energy:Plancherel} can be calculated in any transverse plane after some manipulations:
\begin{equation}\label{eq:appendix:proofenergy:energy:Plancherel:integrand}
\begin{split}
& \hat{\psi}_{ E_y} \bar{\hat{\psi}}_{B_z} - \hat{\psi}_{E_z} \bar{\hat{\psi}}_{B_y}  = \\
& \left[ 1 -\left( \frac{1-\mathscr{P}}{1+\mathscr{P}}  \right)^2  \right]  \left(   \hat{\psi}^{(0)}_{ E_y} \bar{\hat{\psi}}^{(0)}_{B_z} - \hat{\psi}_{E_z}^{(0)} \bar{\hat{\psi}}^{(0)}_{B_y} \right).
\end{split}
\end{equation}

Since $0 \leq \mathscr{P}\leq 1$ for forward-propagating waves, then $0\leq (1-\mathscr{P})/(1+\mathscr{P}) \leq 1$ and thus the coefficient in Eq.~\eqref{eq:appendix:proofenergy:energy:Plancherel:integrand} verifies:
\begin{equation}\label{eq:appendix:proofenergy:energy:Plancherel:integrand:coeff}
0 \leq 1 -\left( \frac{1-\mathscr{P}}{1+\mathscr{P}}  \right)^2  \leq 1.
\end{equation}

When substituting Eqs.~\eqref{eq:appendix:proofenergy:energy:Plancherel:integrand}~and~\eqref{eq:appendix:proofenergy:energy:Plancherel:integrand:coeff} into Eq.~\eqref{eq:appendix:proofenergy:energy:Plancherel}, one easily deduces that the total energy is bounded from above by the paraxial-order energy:
\begin{equation}\label{eq:appendix:proofenergy:energy:upper bound}
\begin{split}
&\langle \psi_{E_y} , \psi_{B_z} \rangle - \langle \psi_{E_z} , \psi_{B_y} \rangle \leq \\
&\langle \psi_{E_y}^{(0)} , \psi_{B_z}^{(0)} \rangle - \langle \psi_{E_z}^{(0)} , \psi_{B_y}^{(0)} \rangle,
\end{split}
\end{equation}
because $\hat{\psi}^{(0)}_{ E_y} \bar{\hat{\psi}}^{(0)}_{B_z} - \hat{\psi}_{E_z}^{(0)} \bar{\hat{\psi}}^{(0)}_{B_y} = |C_{0,E_y}^{(0)}|^2+|C_{0,E_z}^{(0)}|^2\geq 0$ everywhere.

Finally, from the bounds~\eqref{eq:appendix:proofenergy:energy:lower bound}~and~\eqref{eq:appendix:proofenergy:energy:upper bound} we conclude that our solution preserves the total energy:
\begin{equation}\label{eq:appendix:proofenergy:energy:conclusion}
\begin{split}
&\langle \psi_{E_y} , \psi_{B_z} \rangle - \langle \psi_{E_z} , \psi_{B_y} \rangle = \\
&\langle \psi_{E_y}^{(0)} , \psi_{B_z}^{(0)} \rangle - \langle \psi_{E_z}^{(0)} , \psi_{B_y}^{(0)} \rangle.
\end{split}
\end{equation}

In conclusion, for the first time to the best of our knowledge, we have demonstrated that, when the terms in the Lax series are computed in the way presented in this paper, the paraxial level sets the total energy and high-order corrections do not modify it. This is in complete agreement with the nature of the wave equation. By observing its solution~\eqref{eq:Illia solver:psi}, the energy is set when prescribing whatever two laser field components in a chosen transverse plane, e.g., in virtue of Eqs.~\eqref{eq:appendix:proofenergy:sol Ey}~and~\eqref{eq:appendix:proofenergy:sol Ez} and Eqs.~\eqref{eq:appendix:proofenergy:sol Ey0}-\eqref{eq:appendix:proofenergy:sol Ez0}. Provided that $0 \leq \mathscr{P}\leq 1$ the propagation phase $ \exp[-2{\rm i}\, \hat{T} (1 - \mathscr{P})\xi/\varepsilon^2] $ in Eq.~\eqref{eq:Illia solver:psi}, whose $\varepsilon$-dependent part is introduced in the Lax series by all the high-order corrections, models the transport of this amount of energy, which remains unchanged through any transverse plane. This is the reason why the $\varepsilon$-dependence of the total energy that comes out in previous works \cite{Salamin2007} is not physical: it reflects the presence of spurious modes that are adding energy artificially.

Note that the fact that the high-order corrections carry no energy is analogous to perturbative expansions of the wavefunction in Quantum Mechanics in some cases \cite{CohenTannoudji1998_I,CohenTannoudji2000_II}. The quantum wavefunction (here analogous to the total energy) is normalized to unity, that can be the same as the lowest-order of its expansion, i.e., the unperturbed wavefunction (here analogous to the paraxial-level energy).

Since computing inverse Fourier transforms far from focal plane may be computationally expensive due to the large transverse-spatial windows involved, in the following Sec.~\ref{sec:Asymptotic behavior far from focal plane}, we shall calculate the leading term of the asymptotic limit of our Lax-series-based analytical solution far from the focal plane and discuss the adequacy of using that limit as boundary condition for FDTD Maxwell solvers instead of the full solutions presented in Sec.~\ref{sec:convergence}.

\section{\label{sec:Asymptotic behavior far from focal plane}Asymptotic behavior far from focal plane}

Let us assume that, following Eqs.~\eqref{eq:splitMaxwell:Ey0}-\eqref{eq:splitMaxwell:Bz0}, the transverse field components at the paraxial order in the position space are:
\begin{equation}\label{eq:asymptotic:Ey j=0}
\psi_{E_y}^{(0)} = \psi^{(0)} ,
\end{equation}
\begin{equation}\label{eq:asymptotic:Ez j=0}
\psi_{E_z}^{(0)} = 0 .
\end{equation}

The paraxial mode $\psi^{(0)}$, assumed to be forward-propagating and hence to have no evanescent components, expands in the limit $\xi \rightarrow \pm \infty$ as:
\begin{equation}\label{eq:asymptotic:limit psi0}
\psi^{(0)} = \frac{1}{\xi^N} \left[ a_0 + \frac{a_1}{\xi} + \frac{a_2}{\xi^2} + \cdots   \right],
\end{equation}
where $N > 0$ is the leading exponent of the asymptotic limit (in general, $N$ is not necessarily an integer), which implies that $a_0 \neq 0$, and all coefficients $a_j = a_j(\upsilon,\zeta,\tau)$ do not depend on $\xi$. Equation~\eqref{eq:asymptotic:limit psi0} verifies the paraxial equation~\eqref{eq:sup_mat_prxl:paraxial yz}:
\begin{equation}\label{eq:asymptotic:paraxial eq}
\begin{split}
&\frac{\partial_\perp^2 a_0}{\xi^N}+ \frac{\partial_\perp^2 a_1-4{\rm i}\,T\,Na_0}{\xi^{N+1}}+\\
& \frac{\partial_\perp^2 a_2-4{\rm i}\,T\,(N+1)a_1}{\xi^{N+2}} + \cdots=0,
\end{split}
\end{equation}
from where we deduce that:
\begin{equation}\label{eq:asymptotic:paraxial eq:aj}
\partial_\perp^{2(j+1)}a_j =0,
\end{equation}
for all $j \geq 0$.

In the scope of this paper, we aim at calculating the asymptotic limit  where $\xi \rightarrow \pm \infty$ of our solution, generically denoted as $\psi^{\infty}$, only at the leading term~${\cal O}(\xi^{-N})$. The particular solutions of high-order corrections for $E_y$ and $E_z$ components, given by Eq.~\eqref{eq:splittingwaveeq:psiperp:P recursion}, vanish by virtue of Eq.~\eqref{eq:asymptotic:paraxial eq:aj} at the leading order~${\cal O}(\xi^{-N})$. Hence, only the homogeneous solutions of high-order corrections may contribute to the limit $\xi \rightarrow \pm \infty$, given by Eqs.~\eqref{eq:splitMaxwell:HEy j>0}~and~\eqref{eq:splitMaxwell:HEz j>0}, at such leading order~${\cal O}(\xi^{-N})$. After some manipulations, the limits for the transverse components are, respectively:
\begin{equation}\label{eq:asymptotic:limit Ey}
\frac{\psi_{E_y}^{\infty}}{\psi^{(0)}} \sim 1 + \frac{1}{a_0}\sum_{j=1}^\infty \varepsilon^{2j} A_{E_y}^{(2j)}, 
\end{equation}
\begin{equation}\label{eq:asymptotic:limit Ez}
\frac{\psi_{E_z}^{\infty}}{\psi^{(0)}} \sim  \phantom{0+} \frac{1}{a_0}\sum_{j=1}^\infty \varepsilon^{2j} A_{E_z}^{(2j)}, 
\end{equation}
where, from Eqs.~\eqref{eq:splitMaxwell:Ey j=1}~and~\eqref{eq:splitMaxwell:Ez j=1} we obtain for $j=1$:
\begin{equation}\label{eq:asymptotic:limit Ey:A2}
 A_{E_y}^{(2)} = \frac{(\partial_\upsilon^2-\partial_\zeta^2)}{16\,T^2}\, a_0 , 
\end{equation}
\begin{equation}\label{eq:asymptotic:limit Ez:A2}
 A_{E_z}^{(2)} = \frac{\partial^2_{\upsilon\zeta}}{8\,T^2}\, a_0 , 
\end{equation}
and, from Eqs.~\eqref{eq:splitMaxwell:HEy j>0},~\eqref{eq:splitMaxwell:HEz j>0}~and~\eqref{eq:asymptotic:paraxial eq:aj} we obtain the following recursive formulae for $j>1$:
\begin{equation}\label{eq:asymptotic:limit Ey:A j>1}
 A_{E_y}^{(2j)} = - \frac{1}{8\,T^2} \left[ \partial_\zeta^2 A_{E_y}^{(2j-2)} -\partial^2_{\upsilon\zeta} A_{E_z}^{(2j-2)}\right] , 
\end{equation}
\begin{equation}\label{eq:asymptotic:limit Ez:A j>1}
 A_{E_z}^{(2j)} = - \frac{1}{8\,T^2} \left[ \partial_\upsilon^2 A_{E_z}^{(2j-2)} -\partial^2_{\upsilon\zeta} A_{E_y}^{(2j-2)}\right].
\end{equation}

From Eqs.~\eqref{eq:asymptotic:limit Ey}-\eqref{eq:asymptotic:limit Ez:A j>1} we see that the leading terms of the limits of $E_y$ and $E_z$ where $\xi \rightarrow \pm \infty$ hinge upon the dominant coefficient $a_0$ in Eq.~\eqref{eq:asymptotic:limit psi0}. The series in Eqs.~\eqref{eq:asymptotic:limit Ey}~and~\eqref{eq:asymptotic:limit Ez} must be truncated at order $\sim {\cal O}(\xi^{-N})$. These limits are first specified below for monochromatic (i.e., $T\rightarrow 1$) Hermite-Gaussian (Appendix~\ref{sec:Hermite-Gaussian beams: Asymptotic behavior far from focal plane}) and Laguerre-Gaussian (Appendix~\ref{sec:Laguerre-Gaussian beams: Asymptotic behavior far from focal plane}) families. Then, these limits are calculated with a time envelope coupled to Hermite-Gaussian (Appendix~\ref{sec:Adding a time envelope to Hermite-Gaussian beams}) and Laguerre-Gaussian beams (Appendix~\ref{sec:Adding a time envelope to Laguerre-Gaussian beams}).

%%%%%%%%%%%%%%%%%%%%%%%%%%%%%%%%%%%%%%%%%%%%%%%%%%%%%%%%%%%
\begin{figure}
\centering
\includegraphics[width=\linewidth]{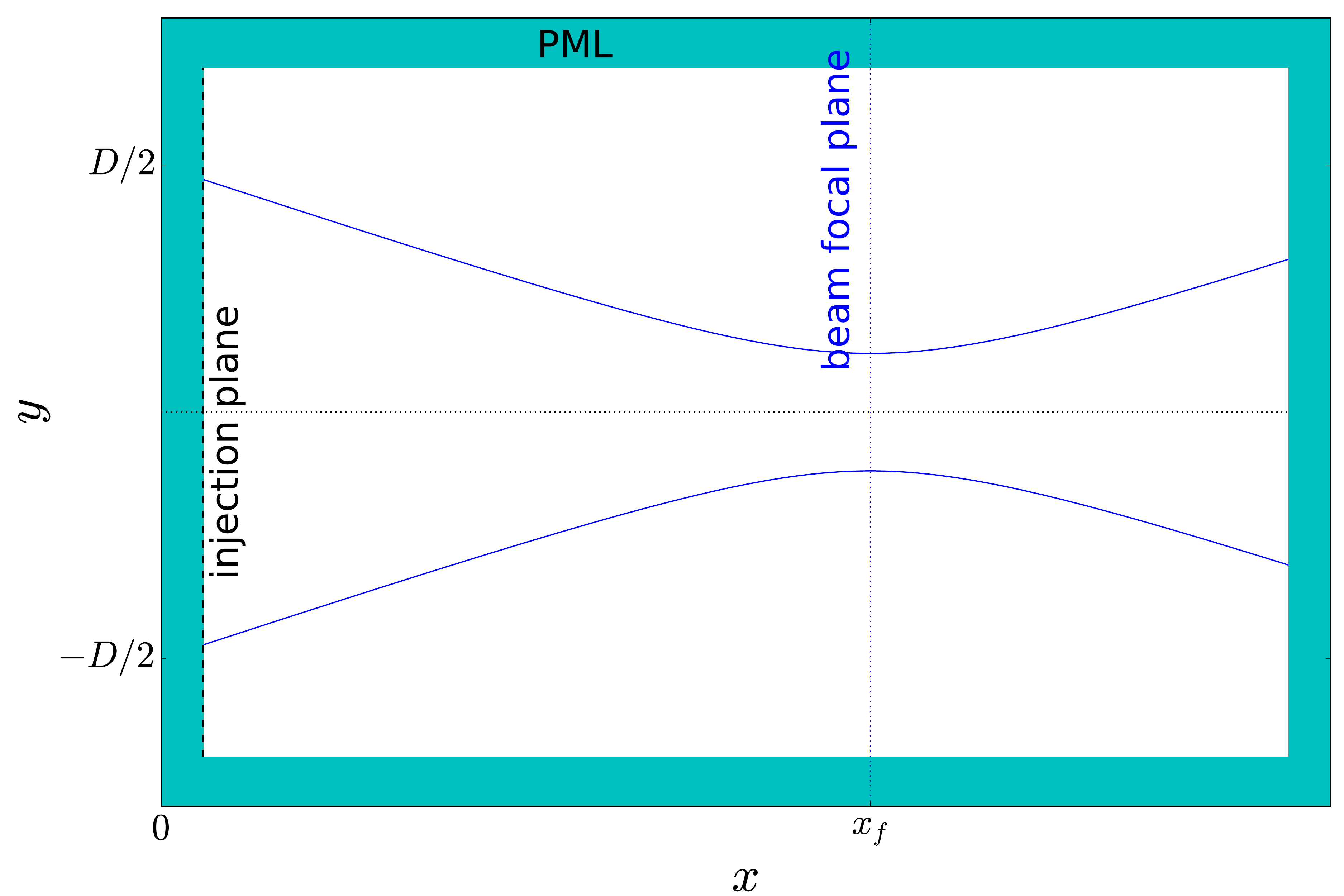}
\caption{\label{fig:domain_ARCTIC}Numerical box in {\sc arctic} (cut at $z=0$). The PML region is colored in cyan. The parameters of the laser beam are defined at $x=0$, namely, the 1/e~diameter $D$ and the numerical aperture. The injection plane for the Total-Field/Scattered-Field technique is placed right after the PML (vertical black dashed line). The beam focal plane (vertical blue dashed line) is at $x=x_f$. Solid blue lines illustrate the 1/e~beam diameter of the corresponding Gaussian pulse.}
\end{figure}
\begin{figure}
\includegraphics[width=\linewidth]{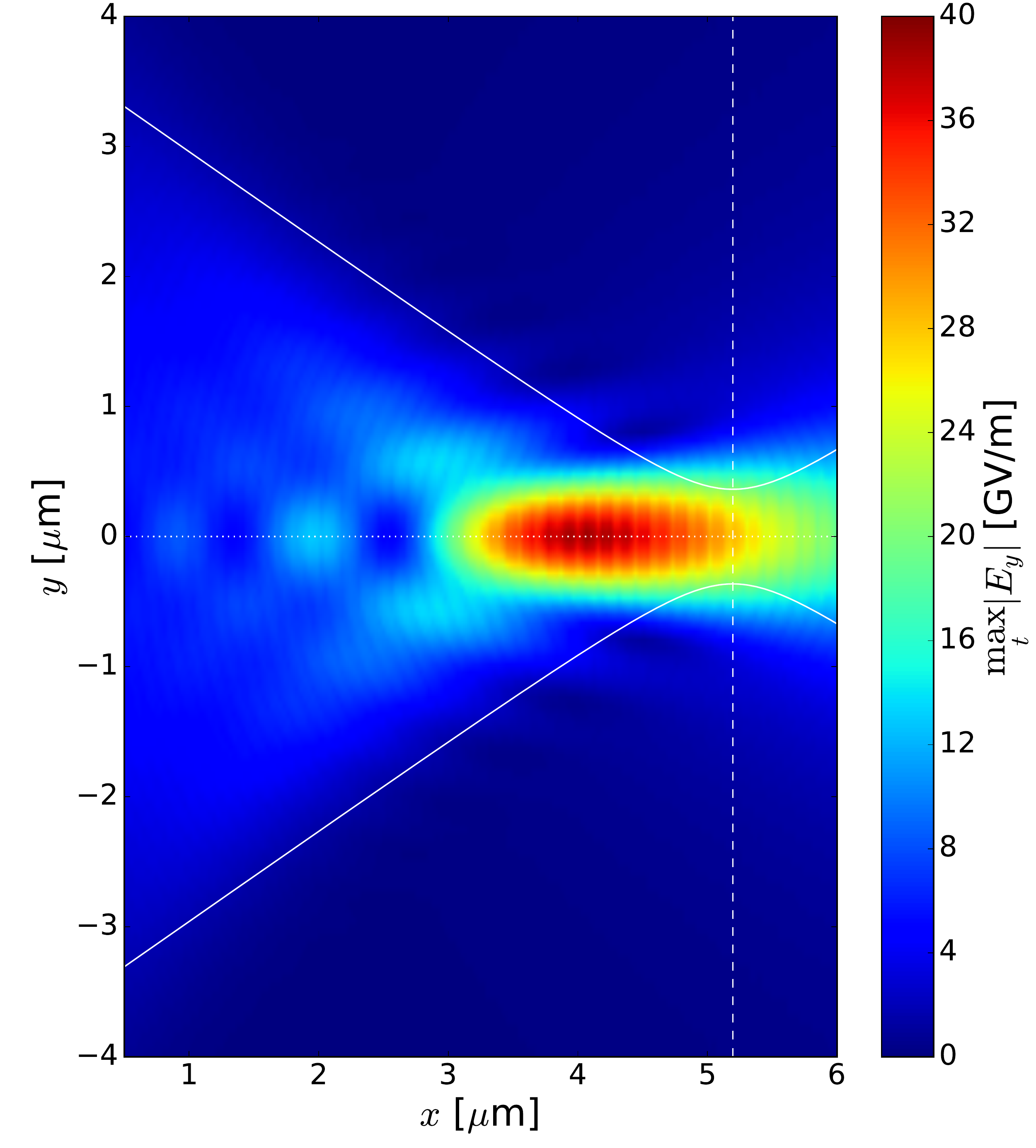}
\caption{\label{fig:ARCTIC_Gauss_maxEy_XY}Maximum value that $|E_y|$ reaches on the $XY$ plane, for a $y$-polarized $0.8$-$\mu$m-wavelength 20-fs-FWHM $36$-nJ Gaussian laser beam prescribed at $x=0$ with ${\rm NA}=0.57$ and $D=7.31$~$\mu$m. The horizontal white dotted line represents the optical axis. The length of the PML layer is $0.32$~$\mu$m along $x$~axis and hence the laser pulse is injected at $x=0.32$~$\mu$m using the paraxial-order term. The vertical white dashed line indicates the beam focal plane position ($x_f=5.20$~$\mu$m) and the white solid lines depict the profile of the Gaussian pulse (paraxial-order term of the Lax series). The leading term of the asymptotic expansion, given by Eq.~\eqref{eq:HEMITExt:asimptote:solution in yzW}, is employed to prescribe the laser pulse.}
\end{figure}
%%%%%%%%%%%%%%%%%%%%%%%%%%%%%%%%%%%%%%%%%%%%%%%%%%%%%%%%%%%

\begin{figure*}
\includegraphics[width=\textwidth]{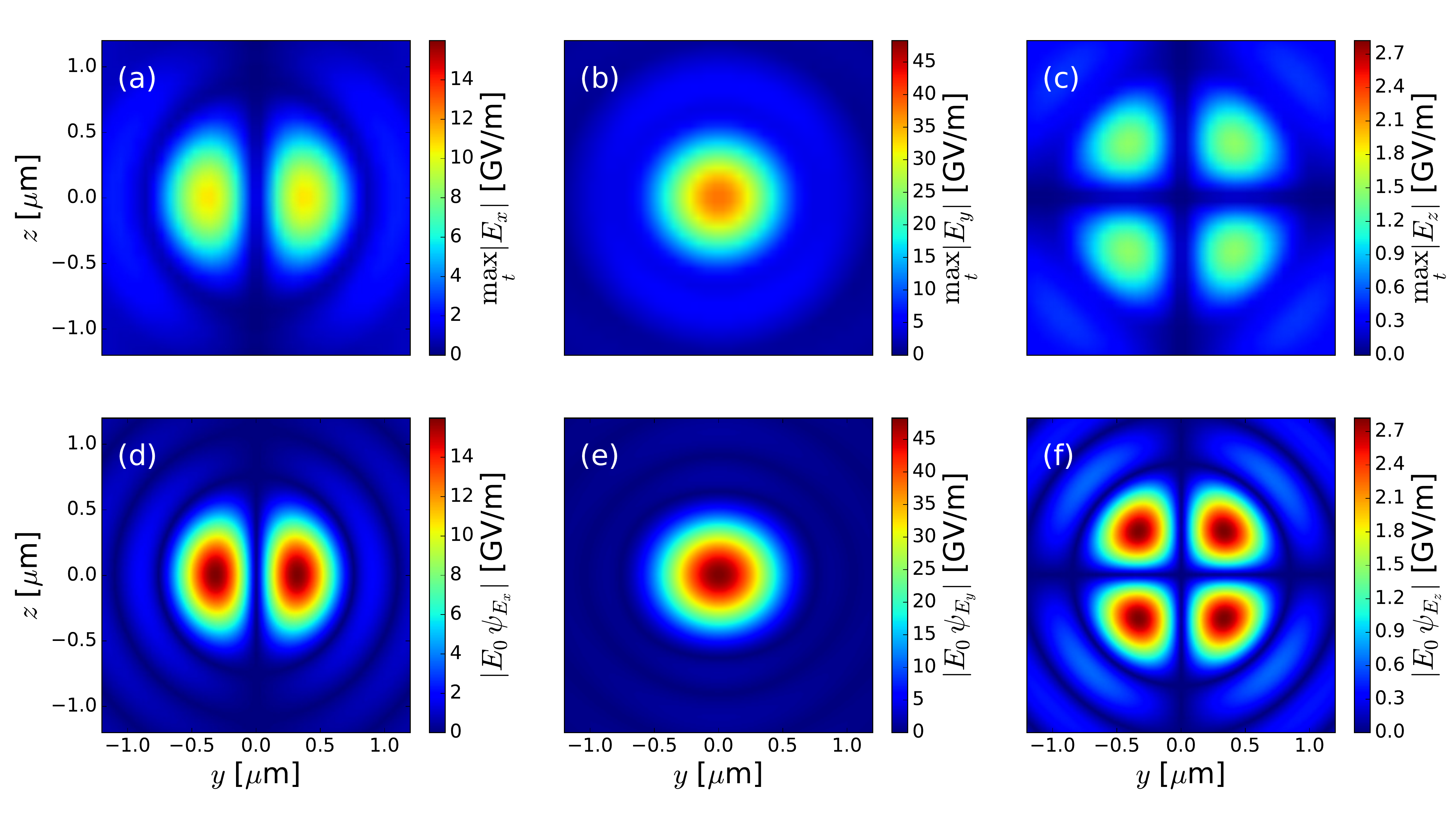}
\caption{\label{fig:ARCTIC_Gauss_cut_focal_plane}Comparison of {\sc arctic}'s results (maximum values that the module of the electric field components attain in the focal plane) with our analytical solution. Results corresponding to Fig.~\ref{fig:ARCTIC_Gauss_maxEy_XY} in the beam focal plane at $x=4.06$~$\mu$m: Maximum values of (a) $|E_x|$, (b) $|E_y|$ and (c) $|E_z|$. Analytical solution in the focal plane at $x_f=5.20$~$\mu$m, calculated from Fourier-backtransformed Eqs.~\eqref{eq:appendix:proofenergy:sol Ey}-\eqref{eq:appendix:proofenergy:sol Bx}, with $T=1$ and by filtering the evanescent modes: (d) $|E_0 \, \psi_{E_x}|$, (e) $|E_0 \, \psi_{E_y}|$ and (f) $|E_0 \, \psi_{E_z}|$.}
\end{figure*}
\begin{figure}
\includegraphics[width=\linewidth]{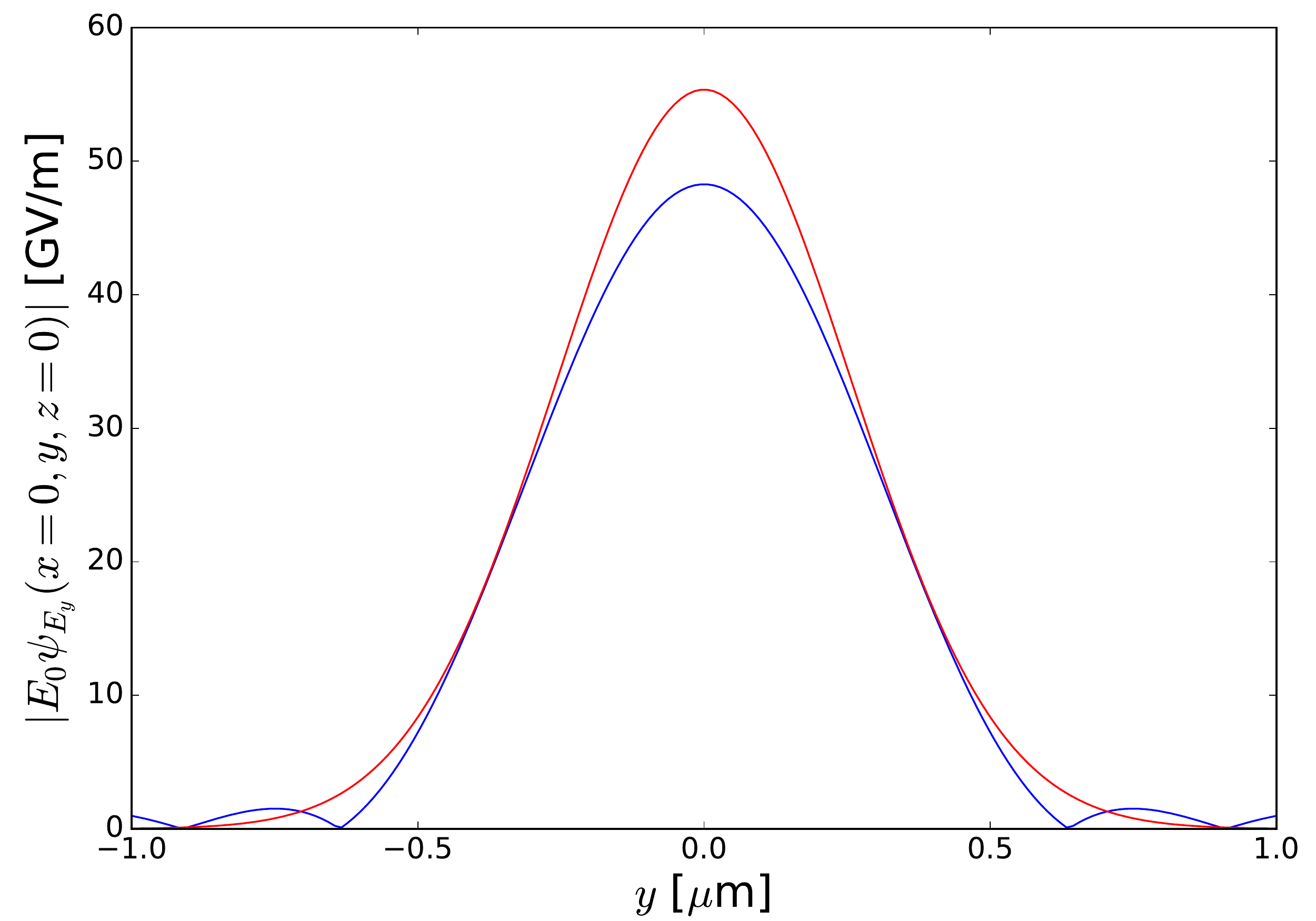}
\caption{\label{fig:ARCTIC_Gauss_profile_cut}Cut along $y$ axis of $|E_0 \, \psi_{E_y}|$ corresponding to Fig.~\ref{fig:ARCTIC_Gauss_cut_focal_plane}(e) (blue solid line). The red solid line accounts for the paraxial-order term of the series.}
\end{figure}

As explained in Sec.~\ref{sec:convergence}, in order to carry out accurate simulations under highly nonparaxial conditions using FDTD Maxwell codes, computing inverse Fourier transforms of Eqs.~\eqref{eq:appendix:proofenergy:sol Ey}-\eqref{eq:appendix:proofenergy:sol Bx} on boundaries is necessary \cite{Thiele2016}. Nevertheless, if the boundaries where fields need to be prescribed are very distant from the focal plane (several Rayleigh lengths), calculating inverse Fourier transforms would demand considerable computational resources because the transverse-spatial window is very large. Alternatively, since our analytical method is capable to link the nonparaxial near fields to the far fields through \eqref{eq:appendix:proofenergy:sol Ey}-\eqref{eq:appendix:proofenergy:sol Bx}, prescribing directly the leading term of the asymptotic limit {\it far enough from the focal plane} appears to be a reasonable simplification that helps us to save a big amount of computational resources in these kind of simulations up to a certain value of $\varepsilon$. Indeed, these leading terms in many cases are simply the paraxial-level term of the Lax series, as shown in Appendices~\ref{sec:Hermite-Gaussian beams: Asymptotic behavior far from focal plane}-\ref{sec:Adding a time envelope to Laguerre-Gaussian beams}, and usually mimic quite well experimental conditions, e.g., a broad beam on a focusing mirror.

We verify our analytical results with three-dimensional (3D) Maxwell-consistent numerical simulations performed using the code {\sc arctic} 
\cite{Thiele2017}. Maxwell equations are discretized by means of Yee scheme \cite{Yee1966}. The simulation domain is delimited by B{\'e}renger's Perfectly-Matched-Layer (PML) absorbing boundary condition \cite{Berenger1994,Berenger1996}. The laser is injected via $E_y$, $E_z$, $B_y$ and $B_z$ components in the transverse plane placed right after the PML according to the Total-Field/Scattered-Field technique \cite{Taflove2005}, as shown in Fig.~\ref{fig:domain_ARCTIC}. This boundary is placed several Rayleigh lengths from the beam focus $x=x_f$.

The origin of the optical axis ($x=0$) is set at the position of the left boundary. The input paraxial-order Gaussian pulse at $x=0$ is characterized by its 1/e~beam diameter $D$ and numerical aperture (NA).  The numerical aperture ($0\leq {\rm NA} \leq 1$) of a Gaussian beam is defined as the sine of its divergence angle. Our Lax series expansion parameter $\varepsilon = (D_0/2)/x_R$, that is, the ratio of the 1/e~beam radius at focus $D_0/2$ and the Rayleigh length $x_R$, represents the tangent of the beam divergence angle. Thus, expressed in terms of $\varepsilon$, the numerical aperture reads:
\begin{equation}\label{eq:NA}
{\rm NA} = \frac{\varepsilon}{\sqrt{1+\varepsilon^2}}.
\end{equation}

The beam focal plane is situated at $x=x_f$:
\begin{equation}\label{eq:xf}
x_f  = \frac{\lambda_0}{\pi \varepsilon^2} \sqrt{\left(  \frac{\pi \varepsilon D}{2  \lambda_0}\right)^2-1}.
\end{equation}

If injecting the leading term of the asymptotic expansion of our solution at $x=0$ (i.e., a simple Gaussian beam) instead of directly injecting the Fourier-backtransformed full solution~\eqref{eq:appendix:proofenergy:sol Ey}-\eqref{eq:appendix:proofenergy:sol Bx}, then the position of the focal plane obtained with the full Maxwell solver may differ from $x_f$ due to simplifying high-order terms where $\xi \rightarrow \pm \infty$ in Eqs.~\eqref{eq:asymptotic:limit Ey}-\eqref{eq:asymptotic:limit Ez}.

We simulate a $y$-polarized Gaussian laser beam at the wavelength $\lambda_0 = 0.8$~$\mu$m coupled, according to Eq.~\eqref{eq:HEMITExt:asimptote:solution in yzW}, with the Gaussian time envelope:
\begin{equation}\label{eq:NUMSIM:Ctau}
C_\tau(\Omega)= \frac{\tau_p}{2\sqrt{\pi}} {\rm e}^{ -\frac{\tau_p^2 \Omega^2}{4} } ={\cal F}_\tau\left[ {\rm e}^{-\frac{\tau^2}{\tau_p^2} }\right],
 \end{equation} 
where $\tau_p=16.99$~fs is the $1/{\rm e}$~duration and thus the Full-Width-at-Half-Maximum (FWHM) duration of the pulse (envelope of intensity) is $20$~fs (i.e., $7.49$~optical cycles). The 1/e~beam diameter at $x=0$ is $D=7.31$~$\mu$m. We take a numerical aperture of ${\rm NA}=0.57$ in the vacuum, which gives $\varepsilon=0.7$ corresponding to strong focusing conditions where the nonparaxial regime is completely established \cite{Salamin2007}. The beam focal plane should be situated at $x_f=5.20$~$\mu$m according to Eq.~\eqref{eq:xf}. Since $x_R=0.52$~$\mu$m, the prescription plane is $x_f/x_R = 10$ Rayleigh lengths far from the beam focal plane. An overall input energy of $36$~nJ is considered (which corresponds to $E_0 = 55.36$~GV/m). The resolution chosen in {\sc arctic} is $\Delta x = 31.8$~nm (25 points per wavelength), $\Delta y = \Delta z =63.7$~nm (13 points per wavelength) and $\Delta t = 84.9$~as (31 points per period). The PML layer is ten cells wide in each direction.

Figure~\ref{fig:ARCTIC_Gauss_maxEy_XY} shows the maximum value of $|E_y|$ over time in the $XY$ plane (i.e., $z=0$). The laser pulse is prescribed according to Eq.~\eqref{eq:HEMITExt:asimptote:solution in yzW}. The temporal inverse Fourier transformed is computed from Eq.~\eqref{eq:sup_mat_Fourier:transverse IFFT tau} using the 64-point Gauss-Legendre quadrature formula in the frequency interval $-10/\tau_p \leq \Omega \leq 10/\tau_p$. The error of 22\% between the position of the beam focal plane given by {\sc arctic} ($x=4.06$~$\mu$m) and the theory ($x_f=5.20$~$\mu$m) is due to the fact that only the leading term of the asymptotic solution is taken into consideration. The previous evaluations have been performed within conditions of very tightly focused pulses ($\varepsilon=0.7$). By decreasing $\varepsilon$ to 0.5, the error drops to roughly 10\% (not shown), which is acceptable. Therefore, it turns out that prescribing the laser fields at a {\it finite} distance implies a contribution of next-to-leading orders of the asymptotic expansion: the smaller $\varepsilon$, the smaller the next-to-leading order contribution.

The maximum values in the focal plane over time of the module of the electric field components are comparable with the module of the corresponding spatial envelopes with $T\rightarrow 1$. These latter values, $|E_0 \psi|$, computed by Fourier-backtransforming Eqs.~\eqref{eq:appendix:proofenergy:sol Ey}-\eqref{eq:appendix:proofenergy:sol Bx} with $\hat{T}=1$, are depicted in Fig.~\ref{fig:ARCTIC_Gauss_cut_focal_plane}(d-f). A 64$\times$64-point Gauss-Legendre quadrature formula in the transverse-wavevector region $\kappa_\perp \leq 2/\varepsilon$ (i.e., the evanescent modes are filtered out) is used to compute the inverse discrete Fourier transforms via Eq.~\eqref{eq:sup_mat_Fourier:transverse IFFT perp}. The peak of $E_y$ predicted by our Lax-series-based solution is $48.26$~GV/m, which is lower than the peak of the paraxial-order term of the series ($E_0 = 55.36$~GV/m), as illustrated in Fig.~\ref{fig:ARCTIC_Gauss_profile_cut}, due to the strong focusing conditions. The cuts in the focal plane of the simulation corresponding to Fig.~\ref{fig:ARCTIC_Gauss_maxEy_XY} are shown in Fig.~\ref{fig:ARCTIC_Gauss_cut_focal_plane}(a-c). One observes that the results of {\sc arctic} and our analytical solution qualitatively agree but the amplitudes are $\sim$20\% smaller (the peak of $E_y$ is $37.80$~GV/m).

%%%%%%%%%%%%%%%%%%%%%%%%%%%%%%%%%%%%%%%%%%%%%%%%%%%%%

\section{\label{sec:Conclusion and outlooks}Conclusion and outlooks}

Both the wave equation and the paraxial equation possess an infinite number of solutions. In this paper, we have demonstrated that from any paraxial solution we can build, in a self-consistent fashion, an exact solution of the wave equation for the six electromagnetic field components, assuming forward-propagating linearly-polarized laser pulses, which, to the first time to best of our knowledge, is consistent with Maxwell equations, conserves the energy transported through transverse planes and preserves the symmetry between the electric and magnetic fields. To do so, we have split, following the procedure by Lax~{\it et al.} \cite{Lax1975} and in the transverse-spatial and temporal Fourier space, both the scalar wave equations applied to each electromagnetic field component and to the Maxwell equations. High-order corrections have been separated in a \emph{homogeneous} solution and a \emph{particular} solution. The particular solution is integrated directly from the wave equation. The homogeneous solution, instead, must be calculated so that the whole set of Maxwell equations is satisfied and the existing symmetry between the electric and magnetic fields is preserved. Only then the total laser energy through transverse planes is conserved. We give simple recursive relations in order to obtain these Maxwell-consistent high-order corrections, which are polynomials on the longitudinal coordinate whose coefficients are paraxial modes related to transverse-spatial and temporal derivatives of the paraxial-order term of the Lax series. The convergence of our solution is demonstrated by giving the limits of the series in the transverse-spatial and temporal Fourier space. These limits are of direct application to accurately prescribe tightly-focused ultrashort laser pulses in Maxwell codes.

Since in experiments fields are usually known far from the focal plane, we have derived the leading term if the asymptotic expansion of the full analytical solution of the Maxwell equations. In the case of a strongly focused 20-fs-FWHM Gaussian laser pulse, numerical simulations confirm the reliability of this asymptotic expression up to an accuracy of 10\%. Further developments for next-to-leading orders are expected to decrease this error.

%%%%%%%%%%%%%%%%%%%%%%%%%%%%%%%%%%%%%%%%%%%%%%%%%%%%%

\begin{acknowledgments}

This research was supported by the project ASTGV (Amplitude Syst{\`e}mes Through-Glass Via) from French DGA (Direction G{\'e}n{\'e}rale de l'Armement) funding, and also by the project ELITAS (ELI Tools for Advanced Simulation) CZ.02.1.01/0.0/0.0/16{\_}013/0001793 from European Regional Development Fund. This work was granted access to the HPC resources  of TGCC under the allocation A0030506129 made by GENCI and the allocation 2017174175 made by PRACE. S.~Skupin acknowledges support by the Qatar National Research Fund (Grant No. NPRP~8-246-1-060). The authors are deeply grateful to the referee for his/her constructive input.

\end{acknowledgments}
%%%%%%%%%%%%%%%%%%%%%%%%%%%%%%%%%%%%%%%%%%%%%%%%%%%%%

\appendix

\section{\label{sec:sup_mat_Fourier}Definition of the transverse-spatial and temporal Fourier transform}

Using the dimensionless coordinates $\xi = x' / x_R$, $\upsilon = 2 y / D_0$, $\zeta = 2 z / D_0$, $\kappa_y  = D_0 k_y /2 $, $\kappa_z = D_0 k_z / 2$, $\tau = \omega_0 t'$ and $\Omega = \omega / \omega_0$ we define the transverse-spatial and temporal Fourier transform of $\psi$, denoted as $\hat{\psi}$, as the combination of the transverse-spatial Fourier transform (${\cal F}_\perp$) and the temporal Fourier transform (${\cal F}_\tau$):
\begin{equation}\label{eq:sup_mat_Fourier:transverse FFT}
\hat{\psi}(\xi,\kappa_y,\kappa_z,\Omega) = {\cal F}_\tau \left[    {\cal F}_\perp \left[ \psi(\xi,\upsilon,\zeta,\tau)  \right] \right],
\end{equation}
\begin{equation}\label{eq:sup_mat_Fourier:transverse IFFT}
\psi(\xi,\upsilon,\zeta,\tau) = {\cal F}_\tau^{-1} \left[    {\cal F}_\perp^{-1} \left[ \hat{\psi}(\xi,\kappa_y,\kappa_z,\Omega)  \right] \right],
\end{equation}
where the transverse-spatial Fourier transform is:
\begin{equation}\label{eq:sup_mat_Fourier:transverse FFT perp}
{\cal F}_\perp \left[ \psi   \right] = \frac{1}{4\pi^2}\iint \psi\,{\rm e}^{-{\rm i}(\kappa_y \upsilon + \kappa_z \zeta)}\, d\upsilon \, d\zeta,
\end{equation}
\begin{equation}\label{eq:sup_mat_Fourier:transverse IFFT perp}
\psi = \iint {\cal F}_\perp \left[ \psi   \right] \,{\rm e}^{{\rm i}(\kappa_y \upsilon + \kappa_z \zeta)}\, d\kappa_y \, d\kappa_z,
\end{equation}
and the temporal Fourier transform is:
\begin{equation}\label{eq:sup_mat_Fourier:transverse FFT tau}
{\cal F}_\tau \left[ \psi   \right] = \frac{1}{2\pi}\int \psi\,{\rm e}^{{\rm i}\Omega \tau}\, d\tau,
\end{equation}
\begin{equation}\label{eq:sup_mat_Fourier:transverse IFFT tau}
\psi = \int {\cal F}_\tau \left[ \psi   \right] \,{\rm e}^{-{\rm i}\Omega \tau}\, d\Omega.
\end{equation}

For monochromatic pulses, the temporal Fourier transform defined in Eq.~\eqref{eq:sup_mat_Fourier:transverse FFT tau} reduces to a multiplication by a Dirac delta function $\delta(\Omega)$ in the temporal Fourier space.

\section{\label{sec:sup_mat_prxl}Solutions of the paraxial equations}

The paraxial equation is:
\begin{equation}\label{eq:sup_mat_prxl:paraxial yz}
(\partial_\upsilon^2+\partial_\zeta^2 + 4 {\rm i} \, T \, \partial_\xi ) \psi=(\partial_\upsilon^2+\partial_\zeta^2 + 4 F^2 \, T \, \partial_F ) \psi = 0,
\end{equation}
where the complex longitudinal variable $F = {\rm i}/({\rm i} - \xi)$ has already been introduced by Salamin \cite{Salamin2007}. By rewriting Eq.~\eqref{eq:sup_mat_prxl:paraxial yz} in the transverse-spatial and temporal Fourier space, we can see that the paraxial solution is of the form:
\begin{equation}\label{eq:sup_mat_prxl:paraxial kykz}
\hat{\psi} = C(\kappa_y, \kappa_z, \Omega) \, {\rm e}^{-{\rm i}\frac{\kappa_\perp^2}{4 \, \hat{T}} \xi},
\end{equation}
where $\kappa_\perp^2=\kappa_y^2+\kappa_z^2$ and $C(\kappa_y, \kappa_z,\Omega)$ is a coefficient independent of $\xi$.

Three families of exact solutions for Eq.~\eqref{eq:sup_mat_prxl:paraxial yz} are known when $T\rightarrow 1$ (i.e., monochromatic pulses): Hermite-Gaussian modes (often called the {\it free-space eigenmodes}), Laguerre-Gaussian modes, and Ince-Gaussian modes \cite{Bandres2004}. Each of these families constitute a countably infinite set of orthogonal paraxial solutions, and they are complete \cite{Agrawal1983}.

\subsection{\label{sec:sup_mat_prxl:HERMITE}Hermite-Gaussian modes}

The Hermite-Gaussian modes are a well-known complete family of orthogonal paraxial solutions:
\begin{equation}\label{eq:sup_mat_prxl:HERMITE:HGnm}
\begin{split}
&\psi^{(HG)}_{n,m} (F,\upsilon,\zeta) = \\
&\sqrt{\frac{(2F-1)^{m+n}}{n! \, m! \,2^{n+m}}}H_n\left(\frac{\sqrt{2}\,F\upsilon}{\sqrt{2F-1}}\right)H_m\left(\frac{\sqrt{2}F\zeta}{\sqrt{2F-1}}\right)  F {\rm e}^{-F \rho^2 },
\end{split}
\end{equation}
where $n$ is the order of the Hermite polynomial $H_n$ along $y$ axis, $m$ is the order along $z$ axis, $\rho^2 = \upsilon^2 + \zeta^2$. Hermite polynomials verify:
\begin{equation}\label{eq:sup_mat_prxl:HERMITE:recursive H}
H_n(x) = 2 x H_{n-1}(x) - 2 (n-1) H_{n-2}(x),
\end{equation}
\begin{equation}\label{eq:sup_mat_prxl:HERMITE:eq H}
H_n''(x) - 2 x H_n'(x) + 2 n H_n(x) = 0,
\end{equation}
where $'$ accounts for the derivative with respect to the variable $x$ and the first two polynomials are $H_0(x) = 1$ and $H_1(x) = 2x$.

Hermite-Gaussian propagation modes are orthogonal between one another, with the inner product defined by Eq.~\eqref{eq:sup_mat_innerprod:definition inner prod mono}:
\begin{equation}\label{eq:sup_mat_prxl:HERMITE:inner product HG}
\begin{split}
& \langle  \psi_{n,m}^{(HG)}, \psi_{p,q}^{(HG)} \rangle = \iint_{-\infty}^{+\infty} \psi_{n,m}^{(HG)} \, \bar{\psi}_{p,q}^{(HG)} \, d\upsilon \, d\zeta =\\
& \langle  \psi_{p,q}^{(HG)}, \psi_{n,m}^{(HG)}  \rangle = \iint_{-\infty}^{+\infty} \psi_{p,q}^{(HG)} \bar{\psi}_{n,m}^{(HG)} \, d\upsilon \, d\zeta =\\
& \frac{\pi}{2}\,\delta_n^p\delta_m^q,
\end{split}
\end{equation}
where $\delta_n^p$ refers to Kronecker delta function and the symbol $\bar{\phantom B}$ denotes the complex conjugate.

In the transverse-spatial Fourier space,the $(n,m)$-order Hermite-Gaussian mode reads:
\begin{equation}\label{eq:sup_mat_prxl:HERMITE:FFT HGnm}
\hat{\psi}^{(HG)}_{n,m} = C^{(HG)}_{n,m}  \, {\rm e}^{-{\rm i}\frac{\kappa_\perp^2}{4}\xi},
\end{equation}
where:
\begin{equation}\label{eq:sup_mat_prxl:HERMITE:C FFT HGnm}
C^{(HG)}_{n,m} = \frac{(-{\rm i})^{n+m}}{4\pi \, \sqrt{n! \, m! \,2^{n+m}}} H_n\left( \frac{\kappa_y}{\sqrt{2}}\right) H_m\left( \frac{\kappa_z}{\sqrt{2}}\right){\rm e}^{-\frac{\kappa_\perp^2}{4}}.
\end{equation}

Transverse derivatives of Hermite-Gaussian modes can be expressed as a linear combination of Hermite-Gaussian modes:
\begin{equation}\label{eq:sup_mat_prxl:HERMITE:iky HG}
{\rm i}\kappa_y \hat{\psi}_{n,m}^{(HG)} = -\sqrt{n+1}  \, \hat{\psi}_{n+1,m}^{(HG)} + \sqrt{n} \, \hat{\psi}_{n-1,m}^{(HG)},
\end{equation}
\begin{equation}\label{eq:sup_mat_prxl:HERMITE:ikz HG}
{\rm i}\kappa_z \hat{\psi}_{n,m}^{(HG)} = -\sqrt{m+1}  \, \hat{\psi}_{n,m+1}^{(HG)} + \sqrt{m} \, \hat{\psi}_{n,m-1}^{(HG)},
\end{equation}
where, by notation convention, $\sqrt{n} \, H_{n-1}(x) =0$ if $n=0$.

\subsection{\label{sec:sup_mat_prxl:LAGUERRE}Laguerre-Gaussian modes}

The Laguerre-Gaussian modes are a well-known complete family of orthogonal paraxial solutions:
\begin{equation}\label{eq:sup_mat_prxl:LAGUERRE:LGpl}
\begin{split}
&\psi^{(LG)}_{p,l}  (F,\upsilon,\zeta)  =\\
&  \frac{(2F-1)^p (\sqrt{2}\, F)^{|l|}}{ \displaystyle  \sqrt{\frac{(p+|l|)!}{p!}}   }  (\upsilon + \sgn(l) \, {\rm i} \zeta )^{|l|} L_p^{|l|}\left( \frac{2\rho^2F^2}{2F-1} \right) F {\rm e}^{-F \rho^2 },
\end{split}
\end{equation}
where $p \geq 0$ is the radial index and $l$ is the azimuthal index (it can be negative, zero or positive integer) of the generalized Laguerre polynomial $L_p^{|l|}$, and $\sgn(l)$ is the sign of $l$, i.e., $\sgn(l)=1$ if $l\geq 0$ and $\sgn(l)=-1$ if $l<0$. Generalized Laguerre polynomials verify:
\begin{equation}\label{eq:sup_mat_prxl:LAGUERRE:recursive L}
\begin{split}
& L_{p}^{|l|}(x) = \\
& \frac{ (2p + |l|-1-x) L_{p-1}^{|l|}(x) - (p+|l|-1) L_{p-2}^{|l|}(x)}{p} ,
\end{split}
\end{equation}
where the first two polynomials are $L_{0}^{|l|}(x) = 1$ and $L_{1}^{|l|}(x) = 1+|l|-x$.

Laguerre-Gaussian propagation modes constitute an orthogonal set:
\begin{equation}\label{eq:sup_mat_prxl:LAGUERRE:inner product LG}
\langle  \psi_{p,l}^{(LG)}, \psi_{q,r}^{(LG)} \rangle = \langle  \psi_{q,r}^{(LG)}, \psi_{p,l}^{(LG)}  \rangle = \frac{\pi}{2}\,\delta_p^q\delta_l^r.
\end{equation}

The Gaussian beam belongs to both Hermite-Gaussian and Laguerre-Gaussian families:
\begin{equation}\label{eq:sup_mat_prxl:LAGUERRE:Gaussian beam}
\phi^{(HG)}_{0,0}=\phi^{(LG)}_{0,0}.
\end{equation}

In the transverse-spatial Fourier space, the $(p,l)$-order Laguerre-Gaussian mode reads:
\begin{equation}\label{eq:sup_mat_prxl:LAGUERRE:FFT LGpl}
\hat{\psi}^{(LG)}_{p,l} = C^{(LG)}_{l,p}  \, {\rm e}^{-{\rm i}\frac{\kappa_\perp^2}{4}\xi},
\end{equation}
where:
\begin{equation}\label{eq:sup_mat_prxl:LAGUERRE:C FFT LGpl}
\begin{split}
& C^{(LG)}_{p,l} = \\
& \frac{(-{\rm i})^{2p+|l|}\, \sqrt{p!}}{4\pi \, \sqrt{2^{|l|}\, (p+|l|)!}  } (\kappa_y + \sgn(l) \, {\rm i} \kappa_z )^{|l|}  L_p^{|l|}\left( \frac{\kappa_\perp^2}{2} \right)       {\rm e}^{-\frac{\kappa_\perp^2}{4}}.
\end{split}
\end{equation}

Transverse derivatives of Laguerre-Gaussian modes can be expressed as a linear combination of Laguerre-Gaussian modes:
\begin{equation}\label{eq:sup_mat_prxl:LAGUERRE:-kp2 LG}
\begin{split}
-\kappa_\perp^2 \hat{\psi}_{p,l}^{(LG)} = & -2(2p +|l| +1)  \, \hat{\psi}_{p,l}^{(LG)} \\
& -2\sqrt{(p+1)(p+1+|l|)}  \, \hat{\psi}_{p+1,l}^{(LG)} \\
& -2\sqrt{p(p+|l|)}  \, \hat{\psi}_{p-1,l}^{(LG)},
\end{split}
\end{equation}
where, by notation convention, $\sqrt{p} \, L_{p-1}^{|l|}(x) =0$ if $p=0$.

\section{\label{sec:sup_mat_innerprod}Laser power and energy transported through a transverse plane and definition of the inner product between spatial envelopes}

The Poynting vector is defined as:
\begin{equation}\label{eq:sup_mat_innerprod:poynting}
{\pmb \Pi} = c^2\varepsilon_0 ({\pmb E} \times \bar{{\pmb B}}),
\end{equation}
where the symbol $\bar{\phantom B}$ denotes the complex conjugate. Its longitudinal component is:
\begin{equation}\label{eq:sup_mat_innerprod:poyntingx}
\Pi_x  = c^2\varepsilon_0 (E_y \bar{B}_z - E_z \bar{B}_y),
\end{equation}
whose integral over the transverse coordinates, calculated by employing Ans{\"a}tze~(\ref{eq:Ansatz:E})~and~(\ref{eq:Ansatz:B}), gives the power flux through the transverse planes:
\begin{equation}\label{eq:sup_mat_innerprod:fluence}
P = \frac{ c \varepsilon_0 E_0^2 D_0^2}{4} \iint_{-\infty}^{+\infty} (\psi_{E_y} \bar{\psi}_{B_z} - \psi_{E_z} \bar{\psi}_{B_y}) \, d\upsilon \, d\zeta .
\end{equation}

Integration of Eq.~\eqref{eq:sup_mat_innerprod:fluence} over time, assuming that there is a time-dependent envelope, gives the total laser energy, which should be the same through any transverse plane:
\begin{equation}\label{eq:sup_mat_innerprod:energy}
U = \frac{1}{\omega_0}  \int_{-\infty}^{+\infty} P \, d\tau.
\end{equation}

The form of the integral in Eq.~\eqref{eq:sup_mat_innerprod:energy} suggests us to define the following inner product of spatial envelopes:
\begin{equation}\label{eq:sup_mat_innerprod:definition inner prod}
\langle a , b \rangle :=  \iiint_{-\infty}^{+\infty} a \bar{b} \, d\upsilon \, d\zeta \, d\tau ,
\end{equation}
which gives us the total energy of the laser pulse:
\begin{equation}\label{eq:sup_mat_innerprod:fluence with inner product}
\frac{4 \omega_0 U} { c \varepsilon_0 E_0^2 D_0^2} =  \langle \psi_{E_y} , \psi_{B_z} \rangle - \langle \psi_{E_z} , \psi_{B_y} \rangle.
\end{equation}

Note that for monochromatic beams (i.e., the envelopes do not depend on time) the inner product is defined as:
\begin{equation}\label{eq:sup_mat_innerprod:definition inner prod mono}
\langle a , b \rangle :=  \iint_{-\infty}^{+\infty} a \bar{b} \, d\upsilon \, d\zeta ,
\end{equation}
and, in this case, $\langle \psi_{E_y} , \psi_{B_z} \rangle - \langle \psi_{E_z} , \psi_{B_y} \rangle$ represents the total power flux through transverse planes:
\begin{equation}\label{eq:sup_mat_innerprod:fluence with inner product mono}
\frac{4 P} { c \varepsilon_0 E_0^2 D_0^2} =  \langle \psi_{E_y} , \psi_{B_z} \rangle - \langle \psi_{E_z} , \psi_{B_y} \rangle.
\end{equation}

Following the definition of the inner product, if $x$ is a scalar (i.e., it does not depend on $\upsilon$ and $\zeta$), we have that:
\begin{equation}\label{eq:sup_mat_innerprod:xab}
\langle x a , b \rangle = x \langle a , b \rangle,
\end{equation}
\begin{equation}\label{eq:sup_mat_innerprod:axb}
\langle a , x b \rangle = \bar{x} \langle a , b \rangle.
\end{equation}

Moreover, it follows from the theory of distributions that odd transverse-coordinate and time derivatives are anticommutative and even transverse-coordinate and time derivatives are commutative. For instance:
\begin{equation}\label{eq:sup_mat_innerprod:dua}
\langle \partial_\upsilon a , a \rangle = - \langle a , \partial_\upsilon a \rangle,
\end{equation}
\begin{equation}\label{eq:sup_mat_innerprod:d2ua}
\langle \partial_\upsilon^2 a , a \rangle =  \langle a , \partial_\upsilon^2 a \rangle,
\end{equation}
provided that $a(\upsilon\rightarrow\pm\infty)=0$ and $\partial_\upsilon a(\upsilon\rightarrow\pm\infty)=0$.

\section{\label{sec:sup_mat_Illia}The exact Maxwell solver in the transverse-spatial Fourier space}

We shall adapt the exact Maxwell solver in transverse-spatial Fourier domain of Ref.~\cite{Thiele2016} to the spatial envelopes given in Ans{\"a}tze~(\ref{eq:Ansatz:E})~and~(\ref{eq:Ansatz:B}). Only the solver for monochromatic laser beams is presented here. To do so, those Ans{\"a}tze are substituted into the Maxwell equations and we obtain the following overdetermined system:
\begin{equation}\label{eq:sup_mat_Illia:system}
\left(  \begin{array}{c c c c c c}
k_x & k_y & k_z & 0 & 0  & 0 \\
0   &   0 & 0 & k_x & k_y & k_z \\
0 & -k_z & k_y & -k_0 & 0 & 0 \\ 
k_z & 0 & -k_x & 0 & -k_0 & 0 \\
k_y & -k_x & 0 & 0 &0 & k_0 \\
k_0 & 0 & 0 & 0 &-k_z & k_y \\
0 & -k_0 & 0 & -k_z & 0 & k_x \\
0 & 0 &k_0 & -k_y & k_x & 0
\end{array}  \right)
\left(  \begin{array}{c}
\hat{\psi}_{E_x} \\
\hat{\psi}_{E_y} \\
\hat{\psi}_{E_z} \\
\hat{\psi}_{B_x} \\
\hat{\psi}_{B_y} \\
\hat{\psi}_{B_z}
\end{array}  \right) =
{\pmb 0},
\end{equation}
where $k_x = \sqrt{k_0^2-k_y^2-k_z^2}$ is the longitudinal component of the wavevector. We only consider forward-propagating modes (i.e., $k_x \geq 0$), and hence we require $\hat{\psi}(x,k_y,k_z)=0$ if $k_y^2+k_z^2>k_0^2$.

The system~(\ref{eq:sup_mat_Illia:system}) has a unique solution if we assume that the two transverse components of the electric field, $E_y$ and $E_z$, are known:
\begin{equation}\label{eq:sup_mat_Illia:system Ex}
\hat{\psi}_{E_x} = -\frac{k_y}{k_x} \hat{\psi}_{E_y} - \frac{k_z}{k_x} \hat{\psi}_{E_z},
\end{equation}
\begin{equation}\label{eq:sup_mat_Illia:system Bx}
\hat{\psi}_{B_x} = -\frac{k_z}{k_0} \hat{\psi}_{E_y} + \frac{k_y}{k_0} \hat{\psi}_{E_z},
\end{equation}
\begin{equation}\label{eq:sup_mat_Illia:system By}
\hat{\psi}_{B_y} = -\frac{k_y k_z}{k_0k_x} \hat{\psi}_{E_y} - \frac{k_0^2-k_y^2}{k_0k_x} \hat{\psi}_{E_z},
\end{equation}
\begin{equation}\label{eq:sup_mat_Illia:system Bz}
\hat{\psi}_{B_z} = \frac{k_0^2-k_z^2}{k_0k_x} \hat{\psi}_{E_y} + \frac{k_y k_z}{k_0k_x} \hat{\psi}_{E_z}.
\end{equation}

The transverse components of the electric field are prescribed in the transverse plane at $x=x_0$ and propagated according the following expression:
\begin{equation}\label{eq:sup_mat_Illia:psiEy}
\hat{\psi}_{E_y}(x,k_y,k_z) = \hat{\psi}_{E_y}(x_0,k_y,k_z) {\rm e}^{-{\rm i} (k_0 - k_x)(x-x_0)},
\end{equation}
\begin{equation}\label{eq:sup_mat_Illia:psiEz}
\hat{\psi}_{E_z}(x,k_y,k_z) = \hat{\psi}_{E_z}(x_0,k_y,k_z) {\rm e}^{-{\rm i} (k_0 - k_x)(x-x_0)},
\end{equation}
which is the exact forward-propagating solution of Eq.~\eqref{eq:psi:xiupsilonzeta:Fourier}.

\section{Examples of asymptotic expansions (leading term)}

\subsection{\label{sec:Hermite-Gaussian beams: Asymptotic behavior far from focal plane}Monochromatic Hermite-Gaussian beams}

If $n$ and $m$ are both even integers, the $(n,m)$-order Hermite-Gaussian mode (see Eq.~\eqref{eq:sup_mat_prxl:HERMITE:HGnm}) behaves asymptotically where $\xi\rightarrow\pm \infty$ like:
\begin{equation}\label{eq:HERMITEGAUSS:asimptote}
\psi^{(HG)}_{n,m} \sim \frac{1}{\xi} \left[ a_0  + {\cal O}\left( \xi^{-1}  \right)\right],
\end{equation}
\begin{equation}\label{eq:HERMITEGAUSS:asimptote a0}
a_0 = - \frac{{\rm i}\, \pi \, \sqrt{n! \, m!}}{2^{\frac{n+m}{2}}\left(\frac{n}{2}\right)!\, \left(\frac{m}{2}\right)!}.
\end{equation}

If $n$ is even and $m$ is odd:
\begin{equation}\label{eq:HERMITEGAUSS:asimptote II}
\psi^{(HG)}_{n,m} \sim \frac{1}{\xi^2} \left[ a_0  + {\cal O}\left( \xi^{-1}  \right)\right],
\end{equation}
\begin{equation}\label{eq:HERMITEGAUSS:asimptote a0 II}
a_0 = - \frac{2\sqrt{2} \, \sqrt{n! \, m!}}{2^{\frac{n+m}{2}}\left(\frac{n}{2}\right)!\, \left(\frac{m-1}{2}\right)!}\,\zeta.
\end{equation}

If $n$ is odd and $m$ is even:
\begin{equation}\label{eq:HERMITEGAUSS:asimptote III}
\psi^{(HG)}_{n,m} \sim \frac{1}{\xi^2} \left[ a_0  + {\cal O}\left( \xi^{-1}  \right)\right],
\end{equation}
\begin{equation}\label{eq:HERMITEGAUSS:asimptote a0 III}
a_0 = - \frac{2\sqrt{2} \, \sqrt{n! \, m!}}{2^{\frac{n+m}{2}}\left(\frac{n-1}{2}\right)!\, \left(\frac{m}{2}\right)!} \, \upsilon.
\end{equation}

If both $n$ and $m$ are odd integers, then the asymptotic expansion is:
\begin{equation}\label{eq:HERMITEGAUSS:asimptote IV}
\psi^{(HG)}_{n,m} \sim \frac{1}{\xi^3} \left[ a_0  + {\cal O}\left( \xi^{-1}  \right)\right],
\end{equation}
\begin{equation}\label{eq:HERMITEGAUSS:asimptote a0 IV}
a_0 = \frac{8\, {\rm i} \, \sqrt{n! \, m!}}{2^{\frac{n+m}{2}}\left(\frac{n-1}{2}\right)!\, \left(\frac{m-1}{2}\right)!} \, \upsilon\zeta.
\end{equation}

Whenever $n$ and $m$ are not simultaneously odd integers, by substituting Eqs.~\eqref{eq:HERMITEGAUSS:asimptote a0},~\eqref{eq:HERMITEGAUSS:asimptote a0 II}~and~\eqref{eq:HERMITEGAUSS:asimptote a0 III} into Eqs.~\eqref{eq:asymptotic:limit Ey:A2}~and~\eqref{eq:asymptotic:limit Ez:A2} one deduces that the paraxial-order term dominates far from the focal plane:
\begin{equation}\label{eq:HERMITEGAUSS:asimptote limit Ey}
\psi_{E_y}^{\infty} \sim  \psi^{(HG)}_{n,m},
\end{equation}
\begin{equation}\label{eq:HERMITEGAUSS:asimptote limit Ez}
\psi_{E_z}^{\infty} \sim 0.
\end{equation}

When both $n$ and $m$ are odd integers, the substitution of Eq.~\eqref{eq:HERMITEGAUSS:asimptote a0 IV} into Eqs.~\eqref{eq:asymptotic:limit Ey:A2}~and~\eqref{eq:asymptotic:limit Ez:A2} yields an extra term~$\sim{\cal O}(\xi^{-3})$ in the asymptotic limit of $E_z$:
\begin{equation}\label{eq:HERMITEGAUSS:asimptote limit Ey II}
\psi_{E_y}^{\infty} \sim  \psi^{(HG)}_{n,m} ,
\end{equation}
\begin{equation}\label{eq:HERMITEGAUSS:asimptote limit Ez II}
\psi_{E_z}^{\infty} \sim  \frac{\varepsilon^2}{8\upsilon\zeta} \, \psi^{(HG)}_{n,m}.
\end{equation}

Following Eq.~\eqref{eq:appendix:proofenergy:energy:conclusion}, it is straightforward to verify that these asymptotic limits contain all the power through the transverse plane of the solution.

\subsection{\label{sec:Laguerre-Gaussian beams: Asymptotic behavior far from focal plane}Monochromatic Laguerre-Gaussian beams}

Laguerre-Gaussian modes (see Eq.~\eqref{eq:sup_mat_prxl:LAGUERRE:LGpl}) behave asymptotically where $\xi\rightarrow\pm \infty$ like:
\begin{equation}\label{eq:LAGUERREGAUSS:asimptote}
\psi^{(LG)}_{p,l} \sim   F^{|l|+1}  \left[ a_0  + {\cal O}\left( \xi^{-1}  \right)\right],
\end{equation}
\begin{equation}\label{eq:LAGUERREGAUSS:asimptote:a0}
a_0 = \alpha_0 \, \rho^{|l|} {\rm e}^{{\rm i}l\phi},
\end{equation}
\begin{equation}\label{eq:LAGUERREGAUSS:asimptote:a0 coeff}
\alpha_0 = (-1)^{p} \sqrt{\frac{2^{|l|} p!}{(p+|l|)!}}L_p^{|l|}(0),
\end{equation}
where $F={\rm i}/({\rm i}-\xi)$, $\rho {\rm e}^{\pm {\rm i}\phi} = \upsilon \pm {\rm i}\zeta$, and, in the cylindrical coordinate system, $\rho = \sqrt{\upsilon^2+\zeta^2}$ represents the radial distance and $\phi$ is the azimuth (such that $\upsilon=\rho \cos\phi$ and $\zeta=\rho\sin\phi$). Note that $L_p^{|l|}(0) \neq 0$ for all $ p \geq 0$ and $l$. After some manipulations, when substituting Eq.~\eqref{eq:LAGUERREGAUSS:asimptote:a0} into Eqs.~\eqref{eq:asymptotic:limit Ey:A2}~and~\eqref{eq:asymptotic:limit Ez:A2} and Eqs.~\eqref{eq:asymptotic:limit Ey:A j>1}~and~\eqref{eq:asymptotic:limit Ez:A j>1}, we have:
\begin{equation}\label{eq:LAGUERREGAUSS:asimptote limit Ey:A2}
A^{(2)}_{E_y} = \frac{|l| (|l|-1) }{8} \alpha_0 \, \left(\upsilon + {\rm i}\sgn(l)\zeta\right)^{|l|-2}  ,
\end{equation}
\begin{equation}\label{eq:LAGUERREGAUSS:asimptote limit Ez:A2}
A^{(2)}_{E_z} = \sgn(l) {\rm i} \, A^{(2)}_{E_y}  ,
\end{equation}
which are zero if $|l|\leq1$, and for $j>1$:
\begin{equation}\label{eq:LAGUERREGAUSS:asimptote limit Ey Ez:A j>1}
A^{(2j)}_{E_y} = A^{(2j)}_{E_z} = 0.
\end{equation}

Therefore, the limits for $E_y$ and $E_z$ are, respectively:
\begin{equation}\label{eq:LAGUERREGAUSS:asimptote limit Ey}
\psi_{E_y}^{\infty} \sim \left[   1 + \frac{\varepsilon^2 |l| (|l|-1) }{8\left(\upsilon + {\rm i}\sgn(l)\zeta\right)^2}    \right] \psi^{(LG)}_{p,l},
\end{equation}
\begin{equation}\label{eq:LAGUERREGAUSS:asimptote limit Ez}
\psi_{E_z}^{\infty} \sim    \frac{{\rm i} \varepsilon^2 l (|l|-1) }{8\left(\upsilon + {\rm i}\sgn(l)\zeta\right)^2}     \psi^{(LG)}_{p,l}.
\end{equation}

\subsection{\label{sec:Adding a time envelope to Hermite-Gaussian beams}Hermite-Gaussian laser pulses}

In the transverse-spatial and temporal Fourier space, we multiply the $(n,m)$-order Hermite-Gaussian mode in the focal plane ($\xi=0$) by a temporal envelope $C_\tau(\Omega)$, in order to prescribe the transverse fields according to Eqs.~\eqref{eq:asymptotic:Ey j=0}~and~\eqref{eq:asymptotic:Ez j=0} with the following paraxial mode:
\begin{equation}\label{eq:HEMITExt:asimptote:solution in Fourier}
\hat{\psi}^{(0)} = C_\tau(\Omega) \, C_{n,m}^{(HG)}(\kappa_y, \kappa_z) \, {\rm e}^{-{\rm i}\frac{\kappa_\perp^2}{\vphantom{\hat{T}^1}4 \hat{T}} \xi},
\end{equation}
which satisfies Eq.~\eqref{eq:sup_mat_prxl:paraxial kykz} and where $C_{n,m}^{(HG)}$ is given by Eq.~\eqref{eq:sup_mat_prxl:HERMITE:C FFT HGnm}. Since by this choice the temporal and transverse-spatial envelopes are separated in the focal plane, the inverse transverse-spatial Fourier transform of Eq.~\eqref{eq:HEMITExt:asimptote:solution in Fourier} is straightforward and thus the paraxial mode in position space reads:
\begin{equation}\label{eq:HEMITExt:asimptote:solution in yzW}
\psi^{(0)} ={\cal F}^{-1}_\tau\left[ C_\tau(\Omega)  \;  \psi^{(HG)}_{n,m} (\tilde{F},\upsilon,\zeta)\right],
\end{equation}
where $\psi^{(HG)}_{n,m}$ is given by Eq.~\eqref{eq:sup_mat_prxl:HERMITE:HGnm} and:
\begin{equation}\label{eq:HEMITExt:asimptote:solution in real yzW:F}
\tilde{F}=\frac{\rm i}{\displaystyle {\rm i}- \xi/\hat{T}}.
\end{equation}

Following Sec.~\ref{sec:Hermite-Gaussian beams: Asymptotic behavior far from focal plane}, whenever $n$ and $m$ are not simultaneously odd integers the asymptotic limits far from the focal plane are:
\begin{equation}\label{eq:HERMITExt:asimptote limit Ey}
\psi_{E_y}^{\infty} \sim {\cal F}^{-1}_\tau\left[ C_\tau(\Omega)  \;  \psi^{(HG)}_{n,m} (\tilde{F},\upsilon,\zeta)\right],
\end{equation}
\begin{equation}\label{eq:HERMITExt:asimptote limit Ez}
\psi_{E_z}^{\infty} \sim 0.
\end{equation}

When both $n$ and $m$ are odd integers:
\begin{equation}\label{eq:HERMITExt:asimptote limit Ey II}
\psi_{E_y}^{\infty} \sim  {\cal F}^{-1}_\tau\left[ C_\tau(\Omega)  \;  \psi^{(HG)}_{n,m} (\tilde{F},\upsilon,\zeta)\right] ,
\end{equation}
\begin{equation}\label{eq:HERMITExt:asimptote limit Ez II}
\psi_{E_z}^{\infty} \sim  \frac{ \varepsilon^2 }{8 T^2 \upsilon\zeta} _, {\cal F}^{-1}_\tau\left[ C_\tau(\Omega)  \;  \psi^{(HG)}_{n,m} (\tilde{F},\upsilon,\zeta)\right] .
\end{equation}

\subsection{\label{sec:Adding a time envelope to Laguerre-Gaussian beams}Laguerre-Gaussian laser pulses}

Analogously to Sec.~\ref{sec:Adding a time envelope to Hermite-Gaussian beams}, we prescribe laser field components based on the following paraxial mode that comes from multiplying a time envelope by a Laguerre-Gaussian mode in the focal plane:
\begin{equation}\label{eq:LAGUERRExt:asimptote:solution in yzW}
\psi^{(0)} ={\cal F}^{-1}_\tau\left[ C_\tau(\Omega)  \;  \psi^{(LG)}_{p,l} (\tilde{F},\upsilon,\zeta)\right],
\end{equation}
where $C_{p,l}^{(LG)}$ is given by Eq.~\eqref{eq:sup_mat_prxl:LAGUERRE:C FFT LGpl} and $\tilde{F}$ is given by Eq.~\eqref{eq:HEMITExt:asimptote:solution in real yzW:F}.

Following Sec.~\ref{sec:Laguerre-Gaussian beams: Asymptotic behavior far from focal plane}, the asymptotic limits for each transverse laser components are, respectively:
\begin{equation}\label{eq:LAGUERRExt:asimptote limit Ey}
\begin{split}
\psi_{E_y}^{\infty} \sim &{\cal F}^{-1}_\tau \left[ C_\tau(\Omega)   \;     \psi^{(LG)}_{p,l} (\tilde{F},\upsilon,\zeta) \right] \times \\
& \left[ 1 + \frac{\varepsilon^2 |l| (|l|-1)}{8 T^2\left(\upsilon + {\rm i}\sgn(l)\zeta\right)^2}    \right],
\end{split}
\end{equation}
\begin{equation}\label{eq:LAGUERRExt:asimptote limit Ez}
\begin{split}
\psi_{E_z}^{\infty} \sim & {\cal F}^{-1}_\tau \left[  C_\tau(\Omega)  \; \psi^{(LG)}_{p,l} (\tilde{F},\upsilon,\zeta) \right] \times \\
&  \frac{{\rm i} \varepsilon^2 l (|l|-1) }{8 T^2\left(\upsilon + {\rm i}\sgn(l)\zeta\right)^2}   .
\end{split}
\end{equation}

%%%%%%%%%%%%%%%%%%%%%%%%%%%%%%%%%%%%%%%%%%%%%%%%%%%%%

\newpage %Just because of unusual number of tables stacked at end
\bibliography{main}% Produces the bibliography via BibTeX.

%%%%%%%%%%%%%%%%%%%%%%%%%%%%%%%%%%%%%%%%%%%%%%%%%%%%%

\end{document}